\documentclass[a4paper, 11pt]{article}
\usepackage[utf8]{inputenc}
\usepackage[T1]{fontenc}
\usepackage{amsmath,amssymb,amsfonts}
\usepackage[title]{appendix}
\usepackage{authblk}
\usepackage{DejaVuSans}
\usepackage{booktabs}
\usepackage{mathtools}
\usepackage{graphicx}
\usepackage[font=small,labelfont=bf]{caption}
\usepackage{subcaption}
\usepackage{microtype}
\usepackage[a4paper,top=2.54cm,bottom=2.54cm,left=1.90cm,right=1.90cm,headsep=1em,includehead,includefoot]{geometry}
\usepackage{url}
\usepackage{hyperref}
\usepackage{cite}

\DeclarePairedDelimiter\abs{\lvert}{\rvert}%
\DeclarePairedDelimiter\norm{\lVert}{\rVert}%

\makeatletter
\let\oldabs\abs
\def\abs{\@ifstar{\oldabs}{\oldabs*}}
\let\oldnorm\norm
\def\norm{\@ifstar{\oldnorm}{\oldnorm*}}
\makeatother

\newcommand{\ra}[1]{\renewcommand{\arraystretch}{#1}}

\providecommand{\keywords}[1]
{
  \small	
  \textbf{\textit{Keywords---}} #1
}

\title{A mechanistic model to assess the effectiveness of test-trace-isolate-and-quarantine under limited capacities}

\author[$1$,$3$,$4$]{Julian Heidecke\thanks{Corresponding author. Email: julian.heidecke@iwr.uni-heidelberg.de}}
\author[$2$]{Jan Fuhrmann}
\author[$1$]{Maria Vittoria Barbarossa}

\affil[$1$]{\small Frankfurt Institute for Advanced Studies,\protect\\ Ruth-Moufang-Straße 1, 60438 Frankfurt, Germany}
\affil[$2$]{Institute of Applied Mathematics, Heidelberg University,\protect\\ Im Neuenheimer Feld 205, 69120 Heidelberg, Germany}
\affil[$3$]{Interdisciplinary Center for Scientific Computing, Heidelberg University,\protect\\ Im Neuenheimer Feld 205, 69120 Heidelberg, Germany}
\affil[$4$]{Heidelberg Institute of Global Health, Heidelberg University,\protect\\ Im Neuenheimer Feld 130.3, 69120 Heidelberg, Germany}

\date{}

\normalsize

\begin{document}
\maketitle
\begin{abstract}
    Diagnostic testing followed by isolation of identified cases with subsequent tracing and quarantine of close contacts -- often referred to as test-trace-isolate-and-quarantine (TTIQ) strategy -- is one of the cornerstone measures of infectious disease control. The COVID-19 pandemic has highlighted that an appropriate response to outbreaks requires us to be aware about the effectiveness of such containment strategies. This can be evaluated using mathematical models. We present a delay differential equation model of TTIQ interventions for infectious disease control. Our model incorporates a detailed mechanistic description of the state-dependent dynamics induced by limited TTIQ capacities. In addition, we account for transmission during the early phase of SARS-CoV-2 infection, including presymptomatic transmission, which may be particularly adverse to a TTIQ based control. Numerical experiments, inspired by the early spread of COVID-19 in Germany, reveal the effectiveness of TTIQ in a scenario where immunity within the population is low and pharmaceutical interventions are absent -- representative of a typical situation during the (re-)emergence of infectious diseases for which therapeutic drugs or vaccines are not yet available. Stability and sensitivity analyses emphasize factors, partially related to the specific disease, which impede or enhance the success of TTIQ. Studying the diminishing effectiveness of TTIQ along simulations of an epidemic wave we highlight consequences for intervention strategies. 
\end{abstract}


\section{Introduction}
	In the absence of effective medication or vaccination, mitigation of an infectious disease relies on so called non-pharmaceutical interventions like mask mandates, hygiene measures, contact restrictions or test-trace-isolate-and-quarantine (TTIQ). These measures aim to reduce epidemiologically relevant contacts (\textit{effective contacts}), viz., those between infectious and susceptible individuals during which the pathogen is successfully transmitted. In contrast to population-wide measures, TTIQ is directly targeted at individuals at risk of being infected. In principle, an effective implementation of TTIQ could allow to ease the need for population-wide contact restrictions and thereby reduce their socio-economic consequences. The TTIQ process begins with searching infected individuals within the population, potentially using some kind of laboratory {\it test} which is typically conducted symptom- or risk-based. Once detected, the so called index cases are asked to {\it isolate} and information concerning their close contacts is collected. For this, public health authorities provide criteria for defining close contacts, depending on available information on the transmissibility of the disease. This definition aims to target contact persons who have potentially been infected by the detected index cases. Based on this information, attempts are then made to reach the close contacts (they are {\it traced}). Successfully approached close contacts are hence asked to self-\textit{quarantine} for a period sufficient to cover their potential incubation and infectious phase. In this way, public health authorities (PHA) aim at catching infected individuals in an early phase, potentially asymptomatic or presymptomatic, in order to prevent further infections. From here on another round of contact tracing can be initiated by tracing contacts of contacts. 
	
	TTIQ is considered a cornerstone of infectious diseases control and was adopted by many countries worldwide to combat the spread of the coronavirus disease 2019 (COVID-19), the disease caused by infection with severe acute respiratory syndrome coronavirus type 2 (SARS-CoV-2). However, its effectiveness is hampered by several factors. Some of these factors are inherent to the TTIQ process:
	\begin{itemize}
	    \item Interviewing index cases about their close contacts and eventually approaching these requires time and results in the so called \textit{tracing delay}.
	    \item Limited capacities of PHA and testing laboratories cause the efficiency and speed of the TTIQ process to decrease if the prevalence of the considered disease, and thus the workload, increases.
	    \item A variety of reasons can prevent individuals from complying with the request to participate in the TTIQ process.
	\end{itemize}  
	Other limitations depend on the nature of the disease in question. For COVID-19 these include:
	\begin{itemize}
	    \item The airborne transmission limits the proportion of identifiable infected contacts of any given detected infectious person, the so called \textit{tracing coverage}, and implies that many potential contacts have to be traced per detected index case. 
	    
	    \item Many infected individuals show weak or unspecific symptoms \cite{Baj2020, RosaMesquita2020}. Moreover, there is a reportedly high proportion of presymptomatic transmissions \cite{RKI_Steckbrief, Ferretti2020, Ganyani2020, He2020}. Therefore, a symptom-based testing strategy will lead to a limited \textit{detection ratio} (the proportion of infectious individuals detected by testing before recovery) and a significant \textit{testing delay} (the combined time between turning infectious, administration of a test, obtaining the test result and starting to isolate) has to be expected. Moreover, tracing the contacts of all symptomatic individuals, which might include many individuals with a common cold or other respiratory disease, is virtually impossible. Therefore, tracing usually has to be limited to the contacts of cases confirmed as infected through a polymerase chain reaction test.

	    \item COVID-19 is associated with a relatively short latency period (time between infection and onset of infectiousness) and infectious period \cite{RKI_Steckbrief,  He2020, Bullard2020, Li2020, Singanayagam2020, Woelfel2020}. When the delays caused by testing and tracing become too large as compared to the timescale of disease progression, a small detection ratio is achieved and the contacts of an identified index case will already have caused infections themselves before they are successfully traced.
	\end{itemize}  
	In light of these limitations, it is important to evaluate the contribution of TTIQ to disease control and to assess how this depends on (i) disease characteristics and (ii) non-disease-specific factors like maintaining low prevalence (to prevent exhaustion of TTIQ capacities), or a high compliance with isolation mandates. For this purpose, mathematical modeling provides a powerful tool. However, incorporating TTIQ into standard mean-field models is challenging due to the individual-based character of contact tracing. Information about the timing, proximity and traceability of contacts of identified index cases has to be lifted from the individual-level to the population-level scale \cite{Mueller2021}. Moreover, in-host processes like the course of infectivity influence the population-level effect of TTIQ. To more readily include such fine grained mechanisms many models use agent- and network-based approaches \cite{Jarvis2021, Kerr2021, Kiss2007, Kojaku2021, Pollmann2021, Quilty2021}, or the corresponding branching process at the onset of the epidemic is studied \cite{Ashcroft2022, Davis2021, Hellewell2020, Klinkenberg2006, Kretzschmar2021, Kretzschmar2020, Kucharski2020, Mueller2016, Mueller2000}. Ordinary (ODE) and delay differential equation (DDE) models \cite{Browne2015, Contreras2021a, Contreras2021, Giordano2020, Lunz2021, Sturniolo2021, Tang2020a, Webb2015} are often less complex, allow for full simulation of the epidemic and are more amenable to analytical investigation. However, to derive quarantine rates in these mean-field settings certain approximations have to be applied. Striking the right balance between an accurate representation of contact tracing and model simplicity is rather challenging. Several proposed models are based on simplified approaches where contact tracing is represented by an increased removal rate of infectious individuals \cite{Giordano2020} or a certain fraction of close contacts is assumed to be instantly quarantined \cite{Tang2020a}. In more rigorous model formulations contact tracing is directly correlated to the identification of index cases \cite{Browne2015, Webb2015}, though these works assume the efficiency of TTIQ to be independent of disease prevalence. In contrast, the model introduced in \cite{Contreras2021} accounts for limited TTIQ capacities assuming perfect efficiency until a certain prevalence/incidence is reached. The later was extended to account for a tracing delay which led to a model based on DDEs \cite{Contreras2021a}. Another approach is presented in \cite{Sturniolo2021} where expressions for the probability that individuals had recent contact with an infectious individual are derived. This probabilistic argument is then used to derive tracing terms that aim at matching the individual- and population-level scales. The model is shown to be in good agreement with an agent-based model for most scenarios. The model in \cite{Lunz2021}, which similarly to \cite{Contreras2021a, Contreras2021} includes a limited testing capacity, focuses on contact heterogeneity by deploying a contact exposure distribution. Based on the idea of digital contact tracing, terms are derived to account for tracing of those contacts that are above a certain exposure threshold. This approach aims at finding an optimal notification threshold that manages to balance disease control and quarantine cost by minimizing unnecessarily quarantined contacts. Among those of the above ODE and DDE models that address COVID-19 \cite{Contreras2021a, Contreras2021, Giordano2020, Lunz2021, Sturniolo2021, Tang2020a}, only the one in \cite{Lunz2021} explicitly accounts for presymptomatic transmissions, a key characteristic of infection with SARS-CoV-2. Models that keep track of the age of infection of infected individuals offer suitable frameworks to incorporate more realistic infectivity profiles and to derive exact contact tracing rates. Such age of infection approaches have been studied previously \cite{Ferretti2020, Pollmann2021, Mueller2016, Mueller2000, Fraser2004, Huo2015, Scarabel2021}, considering also the effect of tracing delays \cite{Ferretti2020, Mueller2016} or of limited capacities \cite{Scarabel2021}, yet again assuming perfect efficiency below a certain incidence threshold.
    
    To aid knowledge and preparedness for future infectious disease outbreaks (COVID-19 or others) we derive a new DDE model that refines and extends approaches previously proposed in the literature. To consider limited laboratory testing capacity, we introduce state-dependent testing rates which are motivated by the considerations in \cite{Barbarossa2021}. Based on mechanistic arguments, we derive contact tracing terms that link the success of index case identification to the quarantine rates achieved by contact tracing. The structure of the resulting terms is similar to those presented in \cite{Contreras2021a}. However, as with the testing terms, our model does not assume the contact tracing efficiency to be constant below a certain prevalence. Instead, the efficiency of testing and tracing is assumed to decrease smoothly with increasing prevalence due to the growing workload. Additionally, we account for the possibility of transmission by individuals in an early (potentially presymptomatic) phase of their infection by introducing a compartment of short duration in which individuals are already infectious but do not yet have an increased chance of being tested unless they are successfully traced (as they cannot yet have developed signs of infection). As a working example we consider a scenario inspired by the spread of COVID-19 in Germany during the second wave in late summer and fall of 2020. This specific situation is of interest for our study as in Germany (and many other European countries) an epidemic wave emerged after a summer of low prevalence, most of the population was still susceptible, pharmaceutical interventions were not widely available, and TTIQ was the main control measure applied in addition to population-wide hygiene and contact interventions. As such a situation is not unusual for (re-)emerging diseases, focusing on this specific example does not restrict the general applicability of our results to other infectious diseases. Our modeling approach sheds light on the effectiveness of TTIQ in this critical early phase of an epidemic, underlines factors that limit the success of this strategy, and demonstrates how its performance is impeded once case numbers begin to increase. Our results highlight implications for intervention strategies and underline conditions for an effective implementation of TTIQ. 
	
\section{Methods}\label{sec:methods}
	The model used in this work extends the known $SEIR$-type model for disease dynamics \cite{Brauer2019}. It comprises equations for compartments representing susceptible individuals ($S$) who can be infected, exposed individuals ($E$) who are infected but not yet infectious, infectious individuals ($U$) who can spread the disease, and removed individuals ($R$) who either recovered or died from the disease. Infections might be confirmed by testing previously undetected infectious individuals. These are assumed to self-isolate ($U\rightarrow I$) and disclose infected contact persons that may be traced and quarantined while still being in the latency period ($E\rightarrow Q_E$) or after turning infectious ($U\rightarrow Q_U$). We consider here solely the tracing and quarantine of individuals who have eventually been infected by the index cases. This means that in the model we neglect the temporary shift to a quarantined compartment for those people who happened to have close contact with the index case, but did not become infected during this event. We assume that quarantined infectious individuals can themselves become confirmed as infectious through testing and are then isolated ($Q_U\rightarrow I$). Removed individuals are not involved in disease transmission, assuming permanent immunity upon recovery -- at least for the short simulation times under consideration. To account for the reportedly high proportion of COVID-19 transmissions occurring prior to the onset of any potential symptoms \cite{RKI_Steckbrief, Ferretti2020, Ganyani2020, He2020}, we separate the infectious phase into two periods. The first period (early infectious phase, all individuals in $U_1,Q_{U_1},I_1$) starts immediately after the exposed phase and marks the onset of infectiousness. We characterize the transition to the second period (late infectious phase, all individuals in $U_2,Q_{U_2},I_2$) by reaching a potential symptom onset. For simplicity, we do not differentiate the individuals in the late infectious phase into those with a symptomatic and those with a completely asymptomatic course of infection (see e.g., \cite{Contreras2021a,Barbarossa2020}) which would introduce additional highly uncertain parameters like the transmission rate and the duration of infectiousness for asymptomatic cases. However, we implicitly account for the presence of asymptomatic or weakly symptomatic cases in the average rate at which individuals in the late infectious stage are detected by being tested.
	
	\begin{table}\centering
		\ra{1.3}
		\captionsetup{width=\textwidth}
		\caption{Dynamic variables of model \eqref{eq:fullsys_presym}.}
		\begin{tabular}{@{}p{1.5cm} p{9.5cm}@{}}\toprule
			Notation & Description\\
			\midrule
			$S$ & susceptible individuals \\
			$E$ & undetected exposed individuals \\
			$Q_E$ & traced and quarantined exposed individuals \\
			$U_{1,2}$ & undetected (early, late) infectious individuals \\
			$Q_{U_{1,2}}$ & traced and quarantined (early, late) infectious individuals \\
			$I_{1,2}$ & confirmed and isolated (early, late) infectious individuals \\
			$R$ & removed individuals \\
			\bottomrule
		\end{tabular}
		\label{tab:var}
	\end{table}
		All the state variables as well as their meanings are summarized in \autoref{tab:var}.  We describe the dynamics of these compartments by the following system of differential equations 
	\begin{figure}
		\centering
		\includegraphics[width=\linewidth]{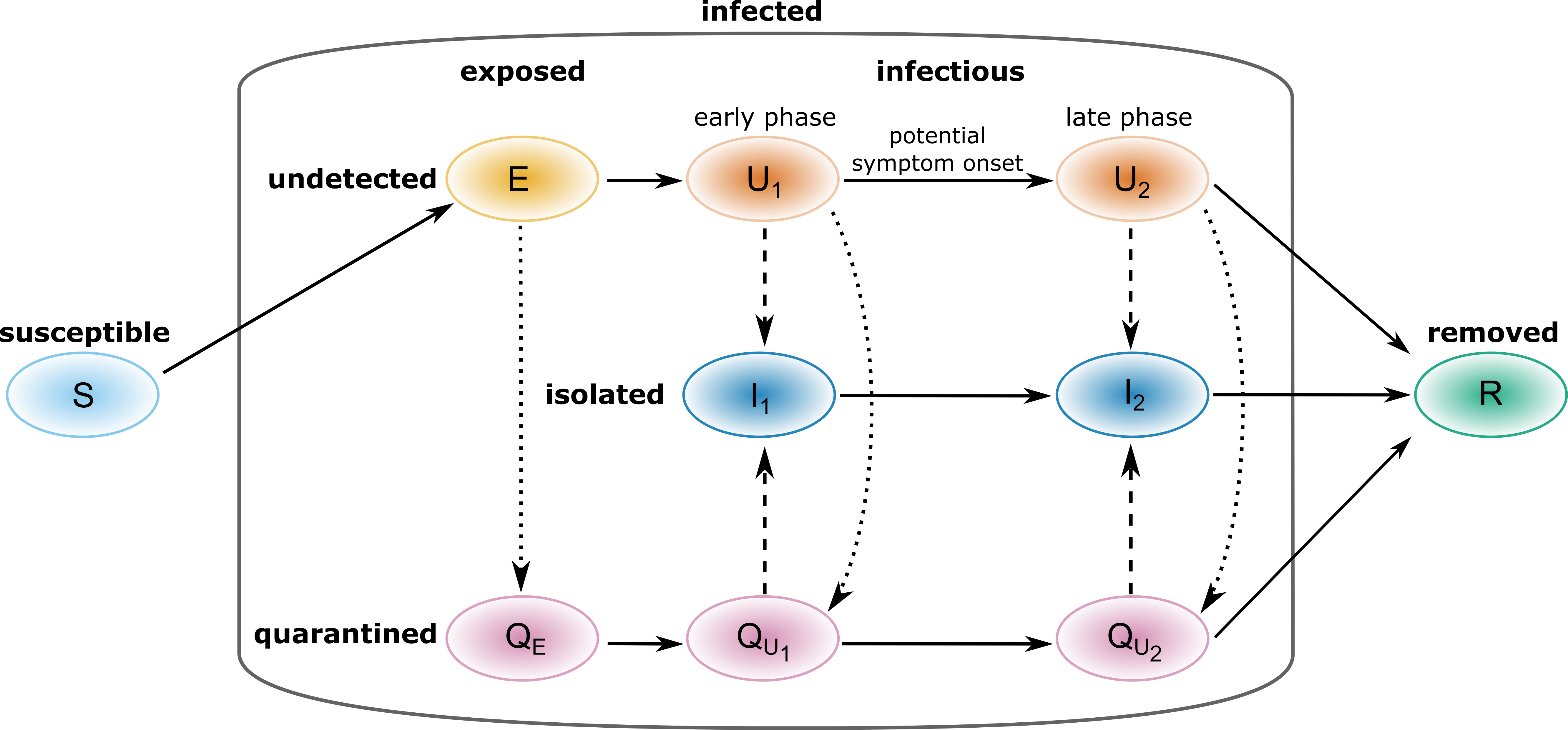}
		\caption{{\bf Flowchart of the transmission model \eqref{eq:fullsys_presym} with TTIQ interventions.} Solid lines indicate state transitions due to disease spread and disease progression. Dashed lines indicate confirmation and isolation of infectious cases due to testing. Dotted lines indicate quarantine of infected contacts due to contact tracing.}
		\label{fig:flowchart_presym}
	\end{figure}
	\begin{equation}\label{eq:fullsys_presym}
		\begin{aligned}
			\dot S &~=~ -\lambda S\\
			\dot E &~=~ \lambda S-\alpha E-Tr_E \\
			\dot Q_E &~=~ Tr_E - \alpha Q_E\\
			\dot U_1 &~=~ \alpha E - \eta_{U_1} U_1 - \gamma_1 U_1 - Tr_{U_1}\\
			\dot Q_{U_1} &~=~ \alpha Q_E + Tr_{U_1} - \eta_{Q_{U_1}} Q_{U_1} - \gamma_1 Q_{U_1}\\
			\dot I_1 &~=~ \eta_{U_1} U_1 + \eta_{Q_{U_1}} Q_{U_1} - \gamma _1 I_1\\
			\dot U_2 &~=~ \gamma_1 U_1 - \eta_{U_2} U_2 - \gamma_2 U_2 - Tr_{U_2}\\
			\dot Q_{U_2} &~=~ \gamma_1 Q_{U_1} + Tr_{U_2} - \eta_{Q_{U_2}} Q_{U_2} - \gamma_2 Q_{U_2}\\
			\dot I_2 &~=~ \eta_{U_2} U_2 + \eta_{Q_{U_2}} Q_{U_2} +\gamma_1 I_1 - \gamma _2 I_2\\
			\dot R &~=~ \gamma_2 (U_2+Q_{U_2}+I_2)
		\end{aligned}
	\end{equation}
	with 
	$$\lambda=\frac{\beta_{U_1} U_1 + \beta_{U_2} U_2 + \beta_{Q_1}Q_{U_1}  + \beta_{Q_2} Q_{U_2} +  \beta_{I_1}I_1  + \beta_{I_2} I_2 }{N},$$ 
	where $N$ denotes the total population size that is assumed to be constant. For simplicity, we neglect demographics, including disease induced deaths, for the simulated period. A sketch of the corresponding transitions between compartments is given in \autoref{fig:flowchart_presym}. The parameters in model \ref{eq:fullsys_presym} are defined as follows: $\beta_X$ denotes the rate of transmission of the disease to susceptibles via contacts with infectious individuals from compartment $X$. We set $\beta_{U_1}=\theta\beta_{U_2}$ for some scaling factor $\theta$ describing the extent of early vs late transmission. Moreover, we assume that undetected infectious individuals, lacking the same knowledge about their risk of being infected, have a higher transmission rate than quarantined infectious individuals ($\beta_{U_i} > \beta_{Q_i},\,\,i\in{1,2}$) who are expected to restrict their contacts to a higher degree. We further assume that quarantined individuals who are not yet confirmed/isolated reduce their contacts less strictly than confirmed cases ($\beta_{Q_i} > \beta_{I_i},\,\,i\in{1,2}$). We express the transmission rate of quarantined and of isolated infectious individuals as 
	\begin{align*}
		\beta_{Q_i}=\rho_Q \beta_{U_i},\,\, \beta_{I_i}=\rho_I \beta_{U_i},\,\, i\in {1,2}.
	\end{align*}
	The parameters $\rho_Q,\rho_{I}\in[0,1]$ describe the fractions of secondary infections caused per unit time by an average quarantined case and by an average isolated case as compared to an average undetected case, respectively. In this sense, $\rho_Q$ and $\rho_I$ capture the efficiency of quarantine and case isolation in reducing transmissions. Motivated by the above discussion we assume $1>\rho_{Q}>\rho_I$. Disease progression is given by the rates $\alpha$, $\gamma_1$ and $\gamma_2$, that is, $1/\alpha$ is the average duration of the latent period and $1/\gamma_{1}$ and $1/\gamma_{2}$ are the average duration of the early and late infectious period, respectively. 
	
	Detection of infectious individuals by testing in compartment $X$ is described by the rates $\eta_X$. Since all the individuals in compartment $U_1$ are without any symptoms, whereas a certain fraction of individuals in $U_2$ show specific symptoms, we assume that the detection rate in compartment $U_2$ is significantly larger $\eta_{U_2}\gg \eta_{U_1}$. Furthermore, it is plausible to assume that the PHA strive to test all traced individuals during their quarantine (independently of whether they show symptoms or not) to identify which of the reported contacts are indeed infected. Thus, we consider the rate at which these individuals are detected to be significantly larger than the corresponding rate for undetected infectious individuals $\eta_Q:=\eta_{Q_{U_1}}=\eta_{Q_{U_2}}\gg \eta_{U_2}$. 
	
	The terms $Tr_X$ describe the rates at which infected individuals in compartment $X$ are quarantined due to contact tracing. We assume that both testing and contact tracing depends on the availability of capacities, so that the corresponding terms are state-dependent. In the following, we describe the derivation of appropriate expressions taking this into account.

	\subsection{Testing terms}
	
	We assume that the per capita detection rates are state-dependent and decrease for larger prevalence of infectious individuals due to finite testing capacity. More precisely, as motivated in \cite{Barbarossa2021}, we set
	\begin{align}\label{eq:detection}
		\eta_{U_1} = \sigma_{U_1} \bar{\eta},\qquad \eta_{U_2} = \sigma_{U_2} \bar{\eta},\qquad \eta_{Q} = \sigma_Q \bar{\eta},
	\end{align}
	where 
	\begin{align*}
		\bar{\eta}:=\frac{\sigma_+}{S+\sum_{X}\sigma_X X + \sigma_-},\qquad X\in \{E,U_1,Q_{U_1},I_1,U_2,Q_{U_2},I_2,R\}.
	\end{align*}
	Here $\sigma_+$ is the maximal number of tests that can be administered and evaluated per unit time, $\sigma_-$ describes the fact that lab capacity cannot be stored and goes to waste unless quickly used, and $\sigma_X$ expresses the relative frequency of getting tested for individuals in compartment $X$ relative to the respective frequency for susceptibles. Setting $\sigma_- = 0$ would correspond to full employment of available lab facilities by administering as many tests as can possibly be processed independent of the prevalence of the disease. As discussed before, it is reasonable to assume that $$\sigma_{Q}\gg\sigma_{U_2}\gg 1 =\sigma_X,\qquad X\in\{E,U_1,I_1,I_2,R\},$$ with $\sigma_Q:=\sigma_{Q_E}=\sigma_{Q_{U_1}}=\sigma_{Q_{U_2}}$, meaning that the chance of being tested is significantly increased for late infectious individuals (because they might show symptoms of the disease) and quarantined individuals (because they were identified as a close contact of an infectious individual). This leads to reasonably high detection rates at low prevalence, but as the prevalence rises, there are more and more individuals who are considered important to receive one of the limited number of tests, the per capita detection rates decrease, and a greater proportion of infections goes undetected. In contrast, a setting where $\sigma_X=1$ for all $X$ would correspond to a scenario where all available tests are used for randomly screening the population and leads to a constant but small detection rate of infections. In comparison to \eqref{eq:detection} a high and constant (that is, independent of the system's state) detection rate in the compartment of late infectious individuals would lead to unrealistically high incidences of confirmed cases as the true prevalence increases. Being aware that the detection rates are not constant but state-dependent we usually omit this dependence in our notation and only specify it when referring to times in the past. 

	\subsection{Contact tracing terms}\label{sec:tracing}

	To make the derivation of the contact tracing terms easier to follow, we describe it here without decomposing the infectious compartments ($U,I,Q$) into an early ($U_1,I_1,Q_{U_1}$) and late ($U_2,I_2,Q_{U_2}$) infectious phase. Thus, in the following we derive the rates at which infected contacts are quarantined from the exposed ($Tr_E$) and from the infectious stage ($Tr_U$). In the almost analogous derivation of the terms ($Tr_E, Tr_{U_1}, Tr_{U_2}$) for the full model \eqref{eq:fullsys_presym}, we take into account that the index cases detected by testing in $U_1$ and $U_2$, respectively, yield different numbers of secondary cases that have been infected on average for different lengths of time before being traced. The quite cumbersome derivation of these terms is included in Appendix A. 
		
	The process of contact tracing is initiated upon detection of an index case by testing. Here we focus on the effect of manual forward contact tracing. This process is based on PHA interviewing the positively tested index case in order to identify contacts they might have infected. Digital contact tracing, which would support this process by using some kind of digital device that keeps track of contacts between individuals and ideally notifies close contacts instantly when index cases are confirmed as infected, as well as backward tracing, which aims to identify the individual that infected the index case, are not considered in this work. The contact tracing process is incorporated into model \eqref{eq:fullsys_presym} based on the following assumptions:
		\begin{itemize}
			\item[(A1)] Contact tracing is triggered by confirmation of infectious cases by testing.
			\item[(A2)] Contact tracing is only triggered by confirmation of previously undetected infectious individuals ($U\rightarrow I$) (\textit{first-order tracing}). More precisely, tracing contacts of contacts (\textit{recursive tracing}) is neglected here.
			\item[(A3)] There is a fixed delay of $\kappa$ units of time between the confirmation of an index case and the start of quarantine for their contacts. This delay represents the duration of the process of interviewing the index case and reaching out to close contacts.
			\item[(A4)] Although we assume them to be imperfectly isolated, index cases only disclose contacts made before their time in isolation.
			\item[(A5)] Only a certain fraction of secondary infections can be identified by PHA via contact tracing.
			\item[(A6)] PHA rely on limited capacities to conduct contact tracing.
		\end{itemize} 
		In reality, all traced close contact persons, including those who did not get infected during their contact to the index case, are asked to quarantine. At the time of contact notification, there is no way to pinpoint the truly infected contacts. Even if contacts quickly receive a laboratory test, some of them may still be in a phase where their infection is not yet detectable. This means that in practice many susceptible individuals would quarantine, but also that additional undetected infected individuals (not necessarily infected by the index case above) could be quarantined by chance. While we account for the tracing effort due to such contact persons, we do not explicitly consider the effects of their quarantine on the outbreak dynamics.
		
\begin{figure}
	    	\centering
	    	\includegraphics[width=0.96\linewidth]{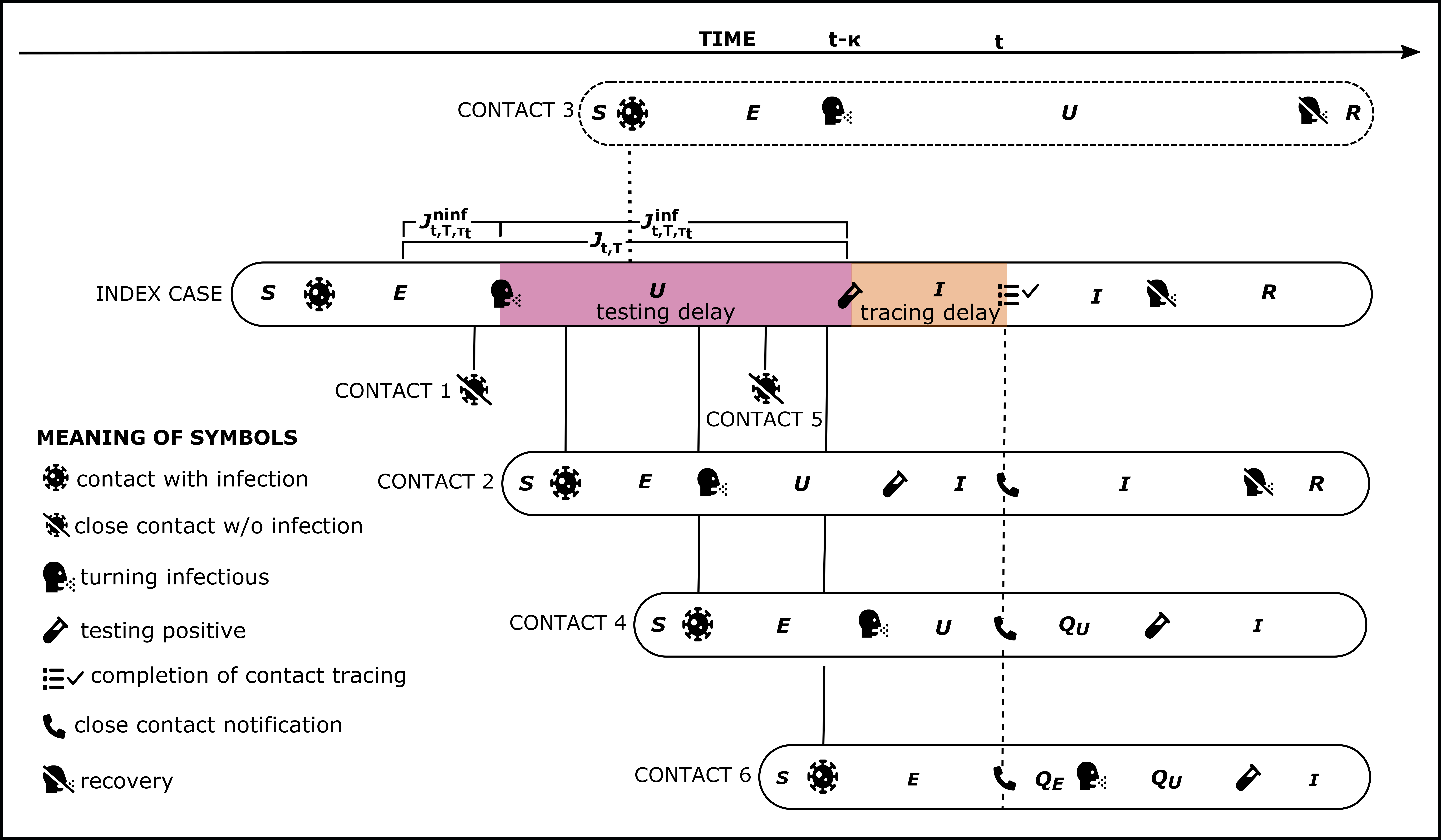}
	    	\caption{{\bf Possible timeline of events for an index case detected by testing at time $t-\kappa$ and contacts made during the tracing interval $J_{t,T}$ for which the index case is asked to disclose close contacts from.} In the depicted example the tracing interval $J_{t,T}$ is longer than the testing delay resulting in Contact $1$ being reported although it could not have been infected since the index case was not yet infectious. Contact $2$ gets infected early during the infectious phase $J^{\text{inf}}_{t,T,}$ of the tracing interval and the time lag resulting from the testing and tracing delay lead to it being already confirmed by testing by the time $t$ of close contact notification. Contact $3$ is an example of an individual that got infected but is missed by the tracing process. This might happen because the contact is not covered by the close contact definition or cannot be recalled by the index case. Contact $4$ and Contact $6$ get infected late in $J^{\text{inf}}_{t,T,}$ shortly before the index case is confirmed as infectious by testing. Therefore, these contacts are less affected by the testing delay, leading to Contact $4$ being undetected but infectious and Contact $6$ even being still latent by the time of being traced $t$. Contact $5$ falls under the definition of being a close contact to the index case and is therefore traced even though no transmission took place. When calculating the tracing efficiency, we account for the tracing effort contacts like $1$ and $5$ generate, however we do not consider the effect of their quarantine on the outbreak dynamics.}
	    	\label{fig:timeline}
	    \end{figure}

		Assumptions (A1)-(A3) imply that the rate at which contacts are quarantined at time $t$ is proportional to $$\eta_U(t-\kappa) U(t-\kappa),$$ that is the rate at which detection of undetected infectious individuals by testing ($U$ to $I$) occurs at time $t-\kappa$. How does this term give rise to the rate at which contacts are quarantined at time $t$? Consider an average index case confirmed as infectious by testing at time $t-\kappa$. We assume that for the interview of this index case the PHA use a certain \textit{tracing window} $T$ that determines the time interval $$J_{t,T}:=[t-\kappa-T,t-\kappa]$$ for which contacts of the index case are registered (\textit{tracing interval}), neglecting that additional contacts made between time $t-\kappa$ and time $t$ could also be reported (see assumption (A4)). Depending on the relationship between the tracing window $T$ and the duration $\tau_t$ for which the average index case has been infectious by the time $t-\kappa$ of being detected, the tracing interval $J_{t,T}$ is composed of two subintervals:
		\begin{itemize}
		    \item the (potentially trivial) part of $J_{t,T}$ during which the index case was not yet infectious
		    \begin{align*}
			J^{\text{ninf}}_{t,T}:=\begin{cases}
				\emptyset, &\quad \text{if}\,\, T\leq\tau_t\\
				[t-\kappa-T,t-\kappa-\tau_t], &\quad \text{if}\,\, T>\tau_t, 
			\end{cases} 
		    \end{align*}
		    
		    \item and the time in $J_{t,T}$ during which the index case has been infectious
		    \begin{align}\label{eq:interv_inf}
	    	J^{\text{inf}}_{t,T}:=\begin{cases}
	    		[t-\kappa-T,t-\kappa], &\quad \text{if}\,\, T\leq\tau_t\\
	    		[t-\kappa-\tau_t,t-\kappa], &\quad \text{if}\,\, T>\tau_t.
	    	\end{cases}
	        \end{align}
		\end{itemize}
		To determine these intervals we need to approximate $\tau_t$. This is discussed under our assumptions on the full model \eqref{eq:fullsys_presym} in \autoref{app:parameter}. We assume that PHA use a certain definition of close contact and ask the index case to report only contacts satisfying this definition. Along with the index case's ability to recall such contacts this determines
		\begin{itemize}
			\item $\widetilde{c}_1$, the reported close contact rate corresponding to the time interval $J^{\text{ninf}}_{t,T}$,
			\item $\widetilde{c}_2$, the reported close contact rate corresponding to the time interval $J^{\text{inf}}_{t,T}$, and
			\item $\widetilde{p}$, the infection probability corresponding to the close contacts made during $J^{\text{inf}}_{t,T}$.  
		\end{itemize} 
	    On the one hand, it is reasonable to assume $\widetilde{c}_2\leq\widetilde{c}_1$, since individuals in $U$ might reduce their contacts upon starting to show symptoms. On the other hand, this is countered by the fact that less recent contacts are harder to recall. Not trying to guess which of these effects is more pronounced, we work with the simplifying assumption $\widetilde{c}:=\widetilde{c}_1=\widetilde{c}_2$. The product $\widetilde{\beta}=\widetilde{p}\,\widetilde{c}$ determines the rate at which the index case produced traceable effective contacts in $J^{\text{inf}}_{t,T}$. Comparing $\widetilde{\beta}$ with $\beta_U$ and $\abs{J^{\text{inf}}_{t,T}}$ with $\tau_t$ gives a proxy for the fraction of infected contacts covered by the contact tracing process (see assumption (A5)). We call the fraction 
	    \begin{align}\label{eq:tr_cov}
	    	\omega(t)=\frac{\abs{J^{\text{inf}}_{t,T}}}{\tau_t}\frac{\widetilde{\beta}}{\beta_U}\leq 1
	    \end{align}
	    the \textit{tracing coverage} at time $t$. Social and hygiene measures obviously influence $\widetilde{c},\,\widetilde{p},\,\widetilde{\beta}$. Moreover, the PHA might change their close contact definition over time or individuals might develop an increased awareness to keep track of their personal contacts to reduce recall issues. We discuss how we handle such measures and time-dependent parameters in Appendix \ref{app:measures}. A possible timeline of events for an index case detected by testing at time $t-\kappa$ whose contacts are quarantined at time $t$ is shown in \autoref{fig:timeline}.
		
\begin{figure}
			\centering
			\includegraphics[width=0.4\linewidth]{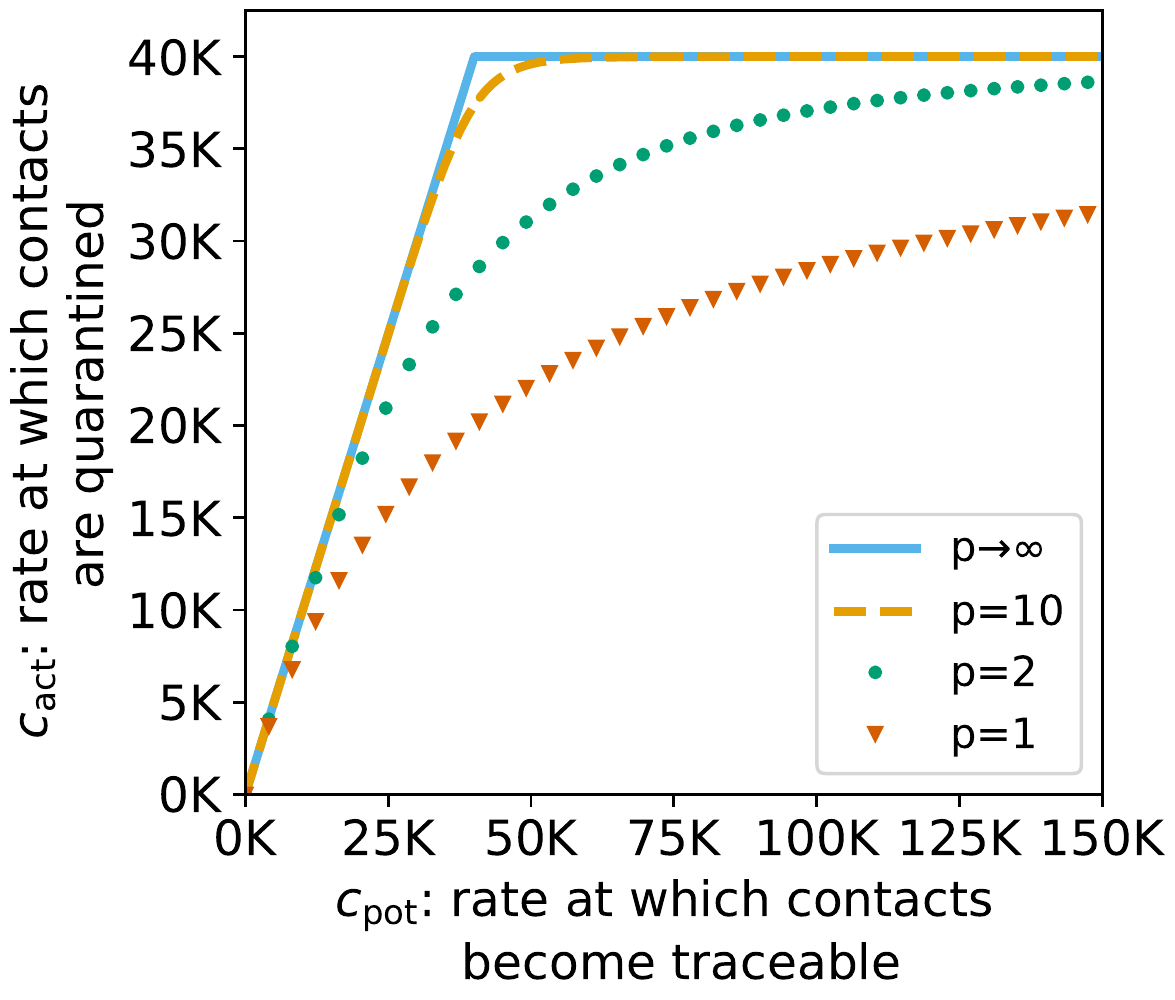}
			\caption{{\bf Illustration of relation \eqref{eq:pNorm} for the choice $\Omega=40\,000$ and multiple values of $p$.} The limit $p\to\infty$ recovers relation \eqref{eq:naive_cap}. Mind the different scaling of the axes.}
			\label{fig:pNorm}
		\end{figure}

		Following the above assumptions we approximate the rate at which contacts become traceable at time $t$ as
		\begin{align}\label{eq:pot}
			c_{\text{pot}}(t)=\abs{J_{t,T}}\,\widetilde{c}\,\eta_U(t-\kappa) U(t-\kappa).
		\end{align} 
		Due to limited resources that can be provided to conduct contact tracing (assumption (A6)) not all traceable contacts necessarily end up being successfully contacted and quarantined. We assume an upper bound $\Omega$ (\textit{tracing capacity}) for the actual rate at which contacts are quarantined due to contact tracing. As a rough approximation we could choose
		\begin{align}\label{eq:naive_cap}
			c_{\text{act}}(t)=\min\left(\Omega,c_{\text{pot}}(t)\right)
		\end{align}
		for the actual rate at which contacts are quarantined at time $t$. However, we suppose the \textit{tracing efficiency}, which we define as the share of successfully traced contacts among all traceable contacts while maintaining the constant tracing delay $\kappa$ and under consideration of the current load on capacities, to be continuously decreasing as the number of traceable contacts increases. That is, we assume that almost perfect efficiency can only be achieved for small rates of traceable contacts. For this reason, we smooth \eqref{eq:naive_cap} by a saturating function of the form
		\begin{align}\label{eq:pNorm}
			c_{\text{act}}(t)=c_{\text{pot}}(t)\underbrace{\frac{\Omega}{\left(c_{\text{pot}}(t)^p+\Omega^p\right)^{1/p}}}_{\substack{=:\varepsilon(t)\text{ tracing efficiency}\\ \text{at time $t$}}},\quad p>0.
		\end{align}
		As shown in \autoref{fig:pNorm}, a larger choice of $p$ (which we call \textit{tracing efficiency constant}) corresponds to a slower initial decrease in tracing efficiency as $c_{\text{pot}}(t)$ rises whereas the limit case $p\to\infty$ gives $$\varepsilon(t)=\frac{\Omega}{\norm{(c_{\text{pot}}(t),\Omega)}_{\infty}},$$ which exactly corresponds to \eqref{eq:naive_cap}. Several factors determine whether the tracing efficiency can be kept high when the prevalence rises (large $p$) or whether it is quickly diminished (small $p$). These include the ease with which additional workforce can be recruited and the efficiency at which this additional workforce (that may have been trained for a different purpose) operates. Moreover, when considering an epidemic outbreak in a large area, the outbreak is usually spatially heterogeneous, with some regions more affected than others. The tracing efficiency can only be kept high in regions with high prevalence when capacity from low prevalence regions can easily be shifted there. If this is not the case (low $p$), then severely affected regions already have low tracing efficiency although the overall prevalence in the considered area might still be relatively low. In a mean-field model like \eqref{eq:fullsys_presym} this is reflected in a fast decreasing tracing efficiency as prevalence rises. As the infection continues to spread, the outbreak is likely to approach states of high prevalence in all regions. As soon as the pool of additionally recruitable workforce is depleted, it is plausible to assume that the considered area makes use of its total capacity and a maximal rate at which contacts can be quarantined is approached.
	
		Only the susceptible close contacts made in the time interval $J^{\text{inf}}_{t,T}$ can have been infected by the index case. The rate at which infected contacts are quarantined at time $t$ is therefore given by
		\begin{align}\label{eq:actinf}
			c_{\text{act}}^{\text{inf}}(t)=\frac{\abs{J^{\text{inf}}_{t,T}}}{\abs{J_{t,T}}} \, \widetilde{p}\,\frac{S(t-\kappa)}{N}\,\,c_{\text{act}}(t).
		\end{align}
		Notice that here we approximate the fraction of susceptible individuals to be constant in the short time interval $J^{\text{inf}}_{t,T,}$. Plugging \eqref{eq:pNorm} into \eqref{eq:actinf}, using \eqref{eq:pot} and $\widetilde{\beta}=\widetilde{p}\,\widetilde{c}$ we get
		\begin{align}\label{eq:generic}
			\underbrace{c_{\text{act}}^{\text{inf}}(t)}_{\substack{\text{actual rate at which} \\ \text{infected contacts are} \\ \text{quarantined at time } t}}=\underbrace{\abs{J^{\text{inf}}_{t,T}}\,\widetilde{\beta}\,\frac{S(t-\kappa)}{N}}_{\substack{\text{secondary infections reported} \\ \text{by an average index case} \\ \text{that was isolated at time } t-\kappa }}\underbrace{\eta_U(t-\kappa) U(t-\kappa)}_{\substack{\text{rate of index case } \\ \text{identification at time } t-\kappa}} \underbrace{\varepsilon(t)}_{\substack{\text{tracing efficiency} \\ \text{at time } t}}.
		\end{align}
		
		Using our proxy \eqref{eq:generic} for the actual rate at which infected contacts are quarantined at time $t$, we can now determine $Tr_E(t)$ and $Tr_U(t)$. Following a similar reasoning as in \cite{Contreras2021a}, we can approximate the fraction of \eqref{eq:generic} originating from the exposed or infectious stage. Let us again consider the average index case detected by testing at time $t-\kappa$. Infection of close contacts reported by this index case took place during $J^{\text{inf}}_{t,T}$ and by the time of being quarantined all of these contacts are infected for at least a duration of $\kappa$ units of time due to the tracing delay. Approximating the infection events produced by our average index case to be uniformly distributed on $J^{\text{inf}}_{t,T}$, we estimate the time an average infected contact spent in the infected chain by the time $t$ of being quarantined as
		\begin{align}\label{eq:time_undetected}
			\widetilde{r}(t)=\kappa+\frac{1}{2}\abs{J^{\text{inf}}_{t,T}}.
		\end{align}
		It should be noticed that the second term on the right-hand side of equation \ref{eq:time_undetected} is a subtle but important addition to the derivation in \cite{Contreras2021a} that takes into account the average time elapsed between infection of close contacts and detection of the index case. A fraction 
		\begin{align*}
			\mu_E(t) := e^{-\alpha \widetilde{r}(t)}
		\end{align*}
		of successfully traced individuals infected by the index case remains in the exposed state ($E\rightarrow Q_E$) after $\widetilde{r}(t)$ units of time in the infected chain. The remaining fraction described by the term $1-\mu_E(t)$ have entered the $U$-compartment but might have been tested themselves or might have recovered by the time of being traced. Therefore, to approximate the fraction of quarantined infected contacts coming from the infectious stage ($U\rightarrow Q_U$) we keep following the first-order kinetics induced by recovery (rate $\gamma$) and testing (rate $\eta_U$) and solve 
		\begin{align}\label{eq:fracU}
			\frac{d\widetilde{U}}{ds} = -(\gamma+\eta_U(t-\kappa))\widetilde{U} + \alpha e^{-\alpha s},\qquad \widetilde{U}(0)=0
		\end{align}
		up to time $s=\widetilde{r}(t)$. Notice that we have assumed the detection rate $\eta_U(t-\kappa)$ to be approximately constant in the short time interval $J^{\text{inf}}_{t,T}$. The solution to the initial value problem \eqref{eq:fracU} evaluated at time $s=\widetilde{r}(t)$ is given by
		\begin{align*}
			\mu_U(t) := \widetilde{U}(\widetilde{r}(t)) = \frac{\alpha}{\eta_U(t-\kappa) + \gamma - \alpha}\left(  e^{-\alpha \widetilde{r}(t)} - e^{- \left(\eta_U(t-\kappa)+\gamma\right) \widetilde{r}(t)}\right).
		\end{align*}
		which continuously extends to $\mu_U(t) = \alpha \widetilde{r}(t)e^{-\alpha \widetilde{r}(t)}$ in the degenerate case $\alpha = \eta_U(t-\kappa) + \gamma$. This motivates us to set
		\begin{align*}
			Tr_E(t)=\mu_E(t)c_{\text{\text{act}}}^{\text{inf}}(t) \quad \text{and} \quad 
			Tr_U(t)=\mu_U(t)c_{\text{\text{act}}}^{\text{inf}}(t).
		\end{align*}
		
	\subsection{Modeling of social and hygiene measures and assessment of TTIQ effectiveness}\label{sec:phistar_phibar}
	    Here, we briefly describe how population-wide social and hygiene measures (e.g., mask mandates, increased hand washing, contact restrictions) are modeled and how we evaluate the effectiveness of TTIQ in controlling disease spread. For a detailed description on how we account for the effect of these measures on the contact tracing terms we refer to Appendix \ref{app:measures}.
	    
	    Population-wide social and hygiene measures are modeled by a time dependent factor $\phi(t)\in[0,1]$, that scales the transmission rates, i.e., 
	    $$\beta_{U_2}(t)=\phi(t)\overline{\beta_{U_2}},$$
	    where $\overline{\beta_{U_2}}$ is the baseline transmission rate of individuals in the undetected late infectious stage corresponding to a phase without any intervention. In case of COVID-19, this can be seen as the transmission rate that would be observed given our contact level before (or in a very early phase during) the first occurrence of the disease. Thus, this baseline transmission rate is associated to the basic reproduction number by
	    \begin{align}\label{eq:R0}
	        \mathcal{R}_0=\frac{\theta\overline{\beta_{U_2}}}{\gamma_1}+\frac{\overline{\beta_{U_2}}}{\gamma_2}.
	    \end{align}
	    The factor $\phi(t)$ can be seen as the level of effective contacts at time $t$ (relative to the "pre-COVID-19" scenario) and leads to the control reproduction number $$\mathcal{R}_{\textbf{c}}:=\phi\mathcal{R}_0.$$ For the evaluation of the effectiveness of TTIQ in preventing an epidemic outbreak, we consider $\phi^*$, the critical level of effective contacts such that the disease-free equilibrium (DFE), $(N,0,...,0)$, is stable for $\phi<\phi^*$ and unstable for $\phi>\phi^*$. In the absence of TTIQ this is clearly given by 
	    \begin{align}\label{eq:basephistar}
	        \phi^*=\frac{1}{\mathcal{R}_0}.
	   \end{align}
	   In the presence of TTIQ we can derive $\phi^*$ from a numerical stability analysis of the DFE (see Appendix \ref{app:stability} for details). The higher the obtained $\phi^*$, the higher the effectiveness of TTIQ and the lower the need for social and hygiene measures. Comparison to the value obtained in the absence of TTIQ \eqref{eq:basephistar} gives a measure of how much population-wide contact restrictions can be relaxed thanks to TTIQ. For a given TTIQ scenario and the corresponding $\phi^*$ we can also interpret 
	   \begin{align}\label{eq:controllable_R}
	        \mathcal{R}_{\text{c}}^{\max}:=\phi^*\mathcal{R}_0    
	   \end{align}
	   as the maximal control reproduction number that can be contained by the given TTIQ effort. Later, we also consider scenarios with high prevalence that operate far from the disease-free state, where stability of the DFE does not offer sufficient information for disease control. In these cases we evaluate the effectiveness of TTIQ based on $\overline{\phi}$, which we define as the critical level of effective contacts that yields a stagnation in the incidence of infected individuals. In other words, for $\phi>\overline{\phi}$ the incidence would further increase and for $\phi<\overline{\phi}$ it would decrease. Clearly, at low prevalence and low immunization we have $\overline{\phi}\approx\phi^*$. 
	   
	   We remark that both $\phi^*$ and $\overline{\phi}$ (and the corresponding values of TTIQ parameters) are threshold values that ensure stability of the DFE or the incidence of infected individuals (i.e., achieve $\mathcal{R}_t\approx 1$). However, in reality the aim is usually to reduce $\mathcal{R}_t$ to a specified value below $1$. Therefore, the derived parameter values should be interpreted as minimally required efforts. In particular, operating at or slightly below $\mathcal{R}_t=1$ would come with the risk of minor changes in some external parameters driving $\mathcal{R}_t$ above $1$, initiating a new wave of exponentially increasing case loads.
	    
\section{Results}\label{sec:results}
     In this section we apply model \eqref{eq:fullsys_presym} to investigate the effectiveness of TTIQ in disease control. As an example of a disease spreading in a population with low immunity, we consider the early spread of COVID-19 in Germany. For this framework, we defined a baseline parameter setting that is motivated in detail in Appendix \ref{app:parameter} and summarized in \autoref{tab:para}. We start by considering the dynamics near the DFE and study the effectiveness of TTIQ in preventing an epidemic outbreak. Examining the sensitivity of our results to parameter changes reveals the disease- and non-disease-specific factors influencing the effectiveness of TTIQ. We then turn to the situation where disease spread is not fully controlled and the effect of TTIQ diminishes as the disease prevalence rises. Revisiting parameter changes reveals implications for intervention strategies.
    \begin{table}\centering
			\ra{1.3}
			\captionsetup{width=.87\textwidth}
			\caption{Parameters of model \eqref{eq:fullsys_presym}. The given values describe our baseline setting which is motivated in Appendix \ref{app:parameter}.}
			\begin{tabular}{@{}p{1.6cm} p{7.6cm}p{1.8cm} p{2.6cm}@{}}\toprule
				Notation & Description & Value & Units \\
				\midrule
				$\overline{\beta_{U_2}}$ & baseline transmission rate of undetected late infectious individuals & $0.33$ & days$^{-1}$people$^{-1}$\\
				$\theta$ & scaling factor for transmission rate in early vs late infectious period & $1.5$ & dimensionless \\
				$(p_Q,p_I)$ & strictness of quarantine and isolation & $(0.2,0.1)$ & dimensionless \\
				$\alpha$ & exposed to early infectious rate & $1/3.5$ & days$^{-1}$  \\
				$\gamma_1$ & early to late infectious rate & $1/2$ & days$^{-1}$ \\
				$\gamma_2$ & late infectious to removed rate & $1/7$ & days$^{-1}$ \\
				$\sigma_+$ & maximal number of tests administered and evaluated per day & $200\,000$ & days$^{-1}$tests \\
				$\sigma_-$ & decay rate of tests & $1.353\,N$ & days$^{-1}$\\
				$(\sigma_{U_2},\sigma_{Q})$ & relative frequency of testing undetected late infectious and traced individuals compared to susceptibles & $(93,300)$ & people$^{-1}$days$^{-1}$\\
				$T$ & tracing window & $9$ & days \\
				$\overline{\widetilde{c}}$ & baseline reported close contact rate & $0.8$ & days$^{-1}$people$^{-1}$ \\
				$\omega$ & tracing coverage & $0.65$ & dimensionless \\
				$\Omega$ & maximal contact quarantine rate & $40\,000$ & days$^{-1}$people \\
				$\kappa$ & tracing delay & $2$ & days \\
				$p$ & tracing efficiency constant & $2$ & dimensionless \\
				$N$ & population size & $83\,000\, 000$ & people \\ 
				\bottomrule
			\end{tabular}
			\label{tab:para}
	\end{table}
	\subsubsection*{Effectiveness of TTIQ in preventing an outbreak}
\begin{figure}
		\centering
		\includegraphics[width=0.75\linewidth]{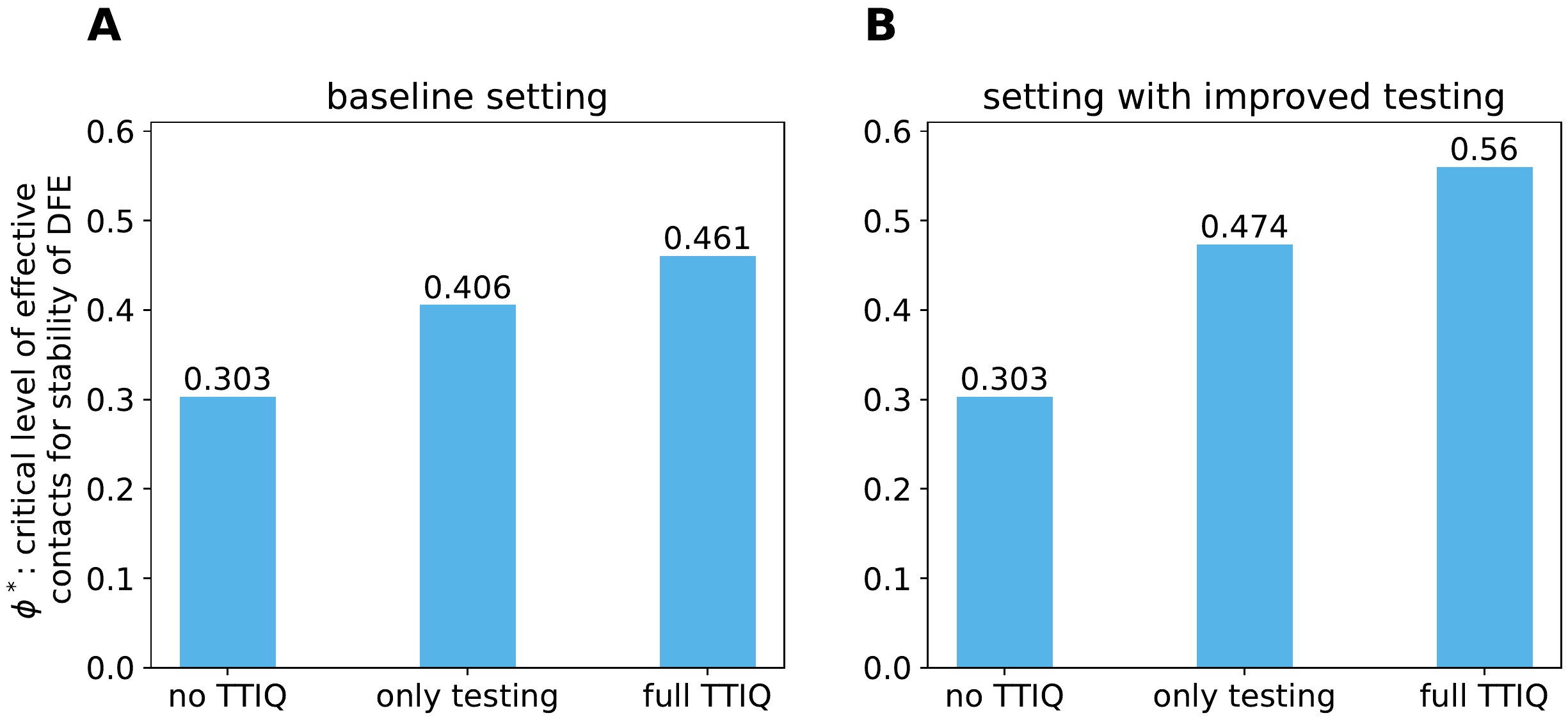}
		\caption{{\bf Critical level $\phi^*$ of effective contacts for the stability of the DFE.} Here we compare a scenario without any TTIQ (no TTIQ), a scenario where only testing is conducted (only testing) and a scenario where testing and tracing is carried out (full TTIQ). This is shown for \textbf{A} the baseline parameter setting given in \autoref{tab:para} and \textbf{B} a setting with improved testing such that $\sigma_{U_2}=185$.}
		\label{fig:noTTI_vs_testing_vs_fullTTI}
	\end{figure}

	We compared values of the critical level $\phi^*$ of effective contacts for the stability of the DFE (explained in Section \ref{sec:phistar_phibar}) for (i) a scenario without TTIQ, (ii) a scenario in which only testing is performed, and (iii) a scenario in which both testing and contact tracing are performed ( \autoref{fig:noTTI_vs_testing_vs_fullTTI}A).
	In our baseline setting we assume $\mathcal{R}_0=3.3$, and thus, get $\phi^*\approx0.304$ in the absence of TTIQ from relation \eqref{eq:basephistar}. Adding testing activity controls the disease at higher rates of effective contacts $\phi^*\approx 0.407$. The addition of contact tracing to testing further increases $\phi^*$ to $\phi^*\approx 0.461$. These values for $\phi^*$ correspond to a maximal control reproduction number that can be contained (see relation \eqref{eq:controllable_R}) of at most $\mathcal{R}_{\text{c}}^{\max}\approx 1.34$ when only testing is applied and $\mathcal{R}_{\text{c}}^{\max}\approx 1.52$ when the full TTIQ approach is applied. That means that if the reproduction number would be brought to less than $1.34$ by other measures, testing alone would be sufficient to suppress an outbreak, and if it were lower than $1.52$, the full TTIQ approach could prevent an outbreak -- assuming that the dynamics were still sufficiently close to the DFE. This shows that in our baseline setting (i) the full TTIQ approach allows for an approximately $52\%$ higher effective contact rate, (ii) testing contributes more to disease control than contact tracing, and (iii) even under the application of both testing and contact tracing a quite severe reduction of effective contacts by other measures to $46\%$ of the pre-COVID-19 level is needed. 
	\subsubsection*{Sensitivity to TTIQ parameters}
\begin{figure}
		\centering
		\includegraphics[width=\linewidth]{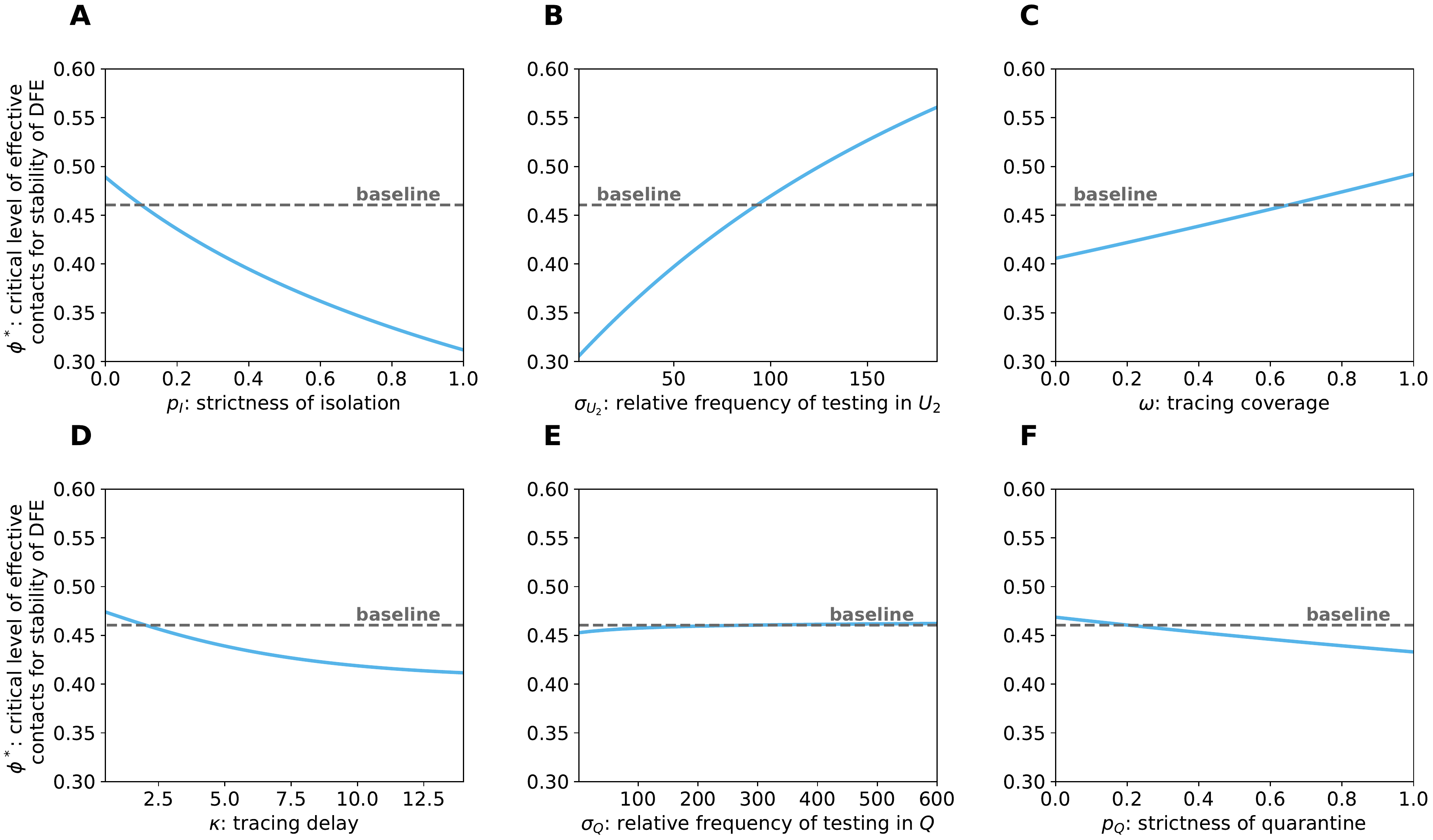}
		\caption{{\bf Critical level $\phi^*$ of effective contacts for the stability of the DFE (solid lines) varying single TTIQ parameters.} For comparison we also show the value of $\phi^*$ derived from our baseline parameter setting given in \autoref{tab:para} (dashed line).}
		\label{fig:crit_red_DFE}
	\end{figure}

\begin{figure}
		\centering
		\includegraphics[width=\linewidth]{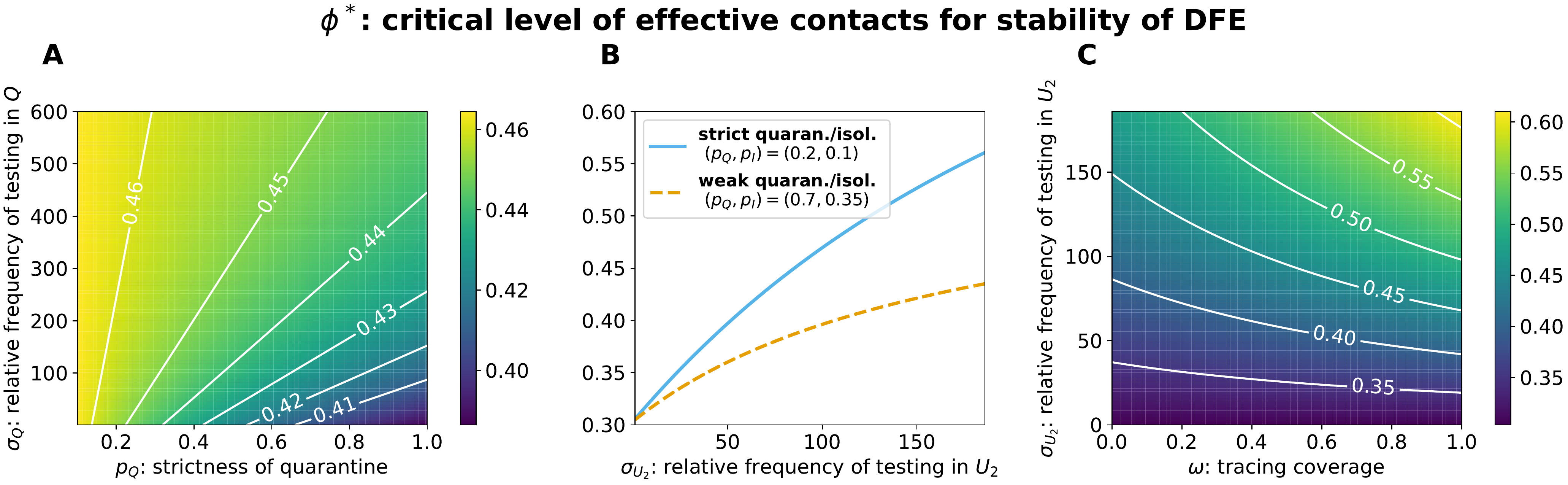} 
		\caption{{\bf Critical level $\phi^*$ of effective contacts for the stability of the DFE varying TTIQ parameters simultaneously.} This is shown for \textbf{A} the strictness of quarantine $p_Q$ and the relative frequency $\sigma_{Q_U}$ of testing traced contacts, \textbf{B} the relative frequency $\sigma_{U_2}$ of testing undetected late infectious individuals assuming either strict quarantine and isolation $(p_Q,p_I)=(0.2,0.1)$ (solid line) or weak quarantine and isolation $(p_Q,p_I)=(0.7,0.35)$ (dashed line) and \textbf{C} the relative frequency $\sigma_{U_2}$ of testing undetected late infectious individuals and the tracing coverage $\omega$.}
		\label{fig:tti_mutually}
	\end{figure}

	To explore alternative parameter settings and determine the most influential TTIQ parameters on $\phi^*$, we first considered single parameters one after the other and varied them in ranges according to \autoref{tab:param_ranges} (\autoref{fig:crit_red_DFE}).
	\begin{table}\centering
		\ra{1.3}
		\captionsetup{width=.77\textwidth}
		\caption{TTIQ parameter ranges considered in the sensitivity analysis.}
		\begin{tabular}{@{}p{1.6cm} p{9.5cm} p{1.2cm}@{}}\toprule
			parameter & meaning & range\\
			\midrule
			$\omega$ & tracing coverage & $[0,1]$ \\
			$\sigma_{U_2}$ & relative frequency of testing undetected late infectious individuals  & $[1,186]$\\
			$\sigma_{Q}$ & relative frequency of testing traced individuals  & $[1,600]$ \\
			$p_Q$ & strictness of quarantine & $[0,1]$\\
			$p_I$ & strictness of isolation & $[0,1]$\\
			$\kappa$ & tracing delay & $[0.5,14]$\\
			\bottomrule
		\end{tabular}
		\label{tab:param_ranges}
	\end{table}
	As shown in \autoref{fig:crit_red_DFE}A, weak isolation of confirmed infectious cases $I$ ($p_I$ close to $1$) leads to $\phi^*\approx0.304$, thus, renders TTIQ completely ineffective. Contact tracing is dependent on prior index case identification and is more effective the shorter the testing delay. This explains why an improved relative frequency $\sigma_{U_2}$ of testing undetected late infectious individuals has the side effect of increasing the effectiveness of contact tracing (compare \autoref{fig:noTTI_vs_testing_vs_fullTTI}A where $\sigma_{U_2}=93$ vs. \autoref{fig:noTTI_vs_testing_vs_fullTTI}B where $\sigma_{U_2}=185$) and why $\sigma_{U_2}$ has a significantly larger impact on $\phi^*$ than the parameters associated with contact tracing $(\omega, \kappa, \sigma_{Q}, p_Q)$ (compare \autoref{fig:crit_red_DFE}B with \autoref{fig:crit_red_DFE}C-E). The tracing coverage $\omega$ appears to be slightly more important than the tracing delay $\kappa$ (compare \autoref{fig:crit_red_DFE}C, \autoref{fig:crit_red_DFE}D). This, however, is mainly due to the baseline parameters we chose, with a relatively short tracing delay $\kappa=2$ and only moderate tracing coverage $\omega=0.65$. Either of these parameters being in an unfavorable range severely limits the effect of any change of the other. In particular, in a setting with long tracing delay but high coverage it is the other way around (not shown here). The strictness of quarantine $p_Q$ as well as the relative frequency $\sigma_{Q}$ at which these individuals are tested appear to be of minor importance (\autoref{fig:crit_red_DFE}E-F). It should be noticed, however, that we chose quite optimistic values for $p_Q$ and $\sigma_{Q}$ in our baseline setting (\autoref{tab:para}). Clearly, $p_Q$ becomes more important when $\sigma_{Q}$ is smaller and $\sigma_{Q}$ gains importance when $p_Q$ is close to unity (\autoref{fig:tti_mutually}A) ($\sigma_Q$ may also be of importance for recursive tracing which is neglected in our equations). Simultaneous improvements of other parameters show synergistic effects. For example, stricter isolation and quarantine reinforces the benefits of improved testing (\autoref{fig:tti_mutually}B), a similar effect is present between improving tracing coverage and improving testing (\autoref{fig:tti_mutually}C), and between improvements of tracing coverage and tracing delay (\autoref{fig:crit_red_dis_char3}A).
    
    To gain insight on the global sensitivity of $\phi^*$ we used latin hypercube sampling on the TTIQ parameter space as given in \autoref{tab:param_ranges} and calculated the corresponding partial rank correlation coefficients (PRCCs). This allows us to assess which TTIQ parameters are most influential on $\phi^*$, even if other parameters are simultaneously perturbed \cite{Marino2008}. We excluded the strictness of quarantine $p_Q$ of traced contacts from this analysis to rule out the unnatural cases $p_Q<p_I$ in which traced (but so far unconfirmed) cases reduce their contacts more strictly than confirmed cases (in these cases increases in $\sigma_{Q}$ would be predicted to have negative impact on $\phi^*$). Instead we allowed $p_I$ and all the remaining parameters given in \autoref{tab:param_ranges} to vary and set 
	$$p_Q=\begin{cases}
		2p_I,\qquad &p_I<0.5\\
		1,\qquad &\text{else}.
	\end{cases}$$
	\begin{figure}
		\centering
		\includegraphics[width=0.4\linewidth]{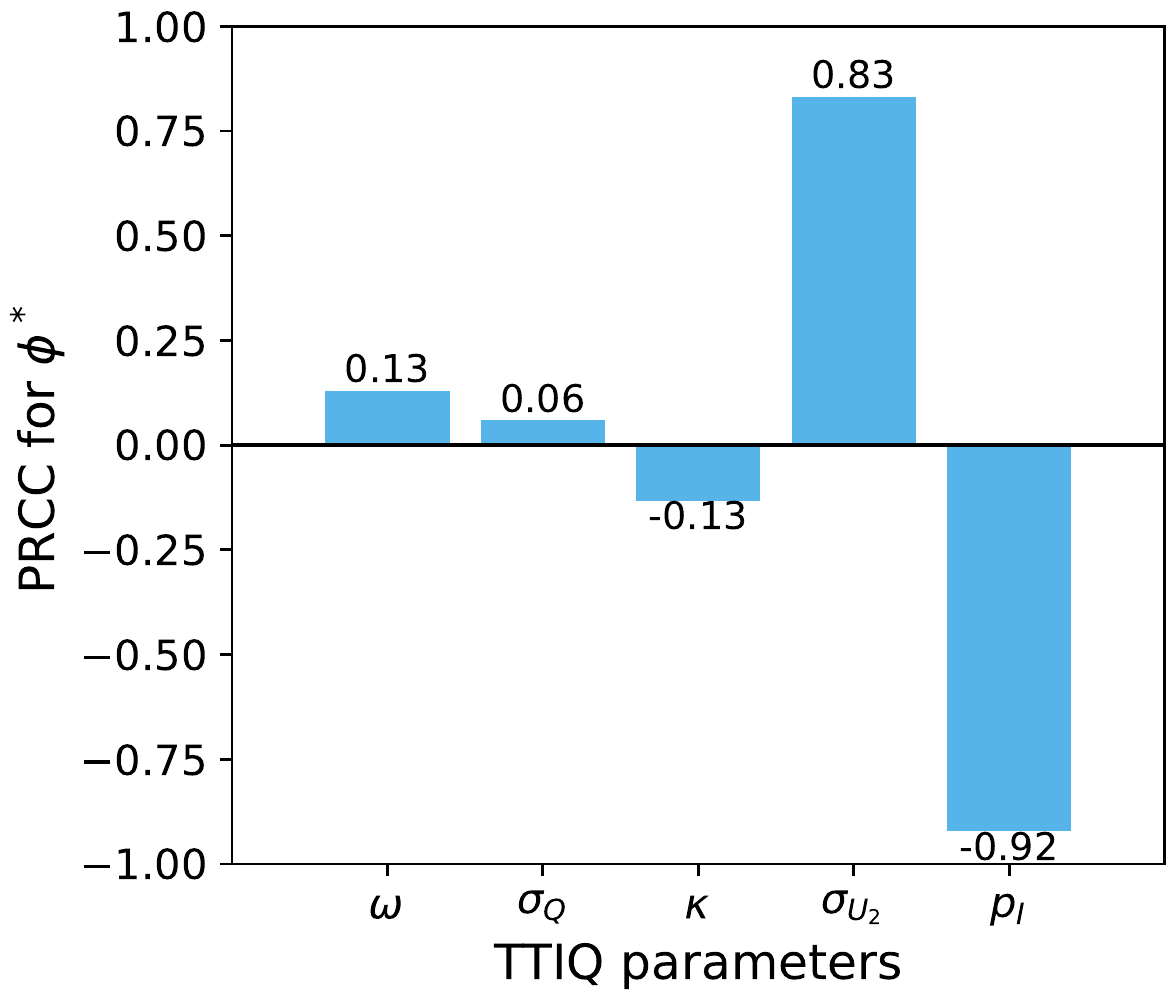}
		\caption{{\bf Global sensitivity of the critical level $\phi^*$ of effective contacts for the stability of the DFE with respect to TTIQ parameters.} The PRCC values where obtained using latin hypercube sampling on the parameter space specified in \autoref{tab:param_ranges}.}
		\label{fig:global_sensitivity}
	\end{figure}
	
	The calculated PRCCs support our observations from the variation of single parameter values. The strictness of isolation $p_I$ of confirmed cases is predicted to be most influential on $\phi^*$ having the largest PRCC (absolute), closely followed by the relative frequency $\sigma_{U_2}$ of testing undetected late infectious individuals which is associated with a significantly larger PRCC than all the parameters corresponding to contact tracing $\omega,\kappa,\sigma_Q$ (\autoref{fig:global_sensitivity}).
    \subsubsection*{Impact of early infectiousness and other disease characteristics}
    We considered variations in parameters describing disease characteristics to investigate the impact of potential uncertainty in our parameter choices and to gain insight on how the effectiveness of TTIQ might be altered for infectious diseases with other characteristics. All parameter changes studied below ensure that $\mathcal{R}_0$ is kept constant, according to formula \eqref{eq:R0}. In particular, in the absence of TTIQ the critical level $\phi^*$ of effective contacts for the stability of the DFE stays unchanged (compare equation \eqref{eq:basephistar}) and all deviations in $\phi^*$ can be attributed to increased or decreased TTIQ effectiveness.
    
    A high incidence of asymptomatic infectious individuals implies a reduced frequency of testing in compartment $U_2$ which we demonstrated to have a major impact on $\phi^*$ (\autoref{fig:crit_red_DFE}B). Therefore, TTIQ has to be expected significantly more effective if overt symptoms occur frequently upon infectious individuals. 
	\begin{figure}
		\centering
        \includegraphics[width=0.7\linewidth]{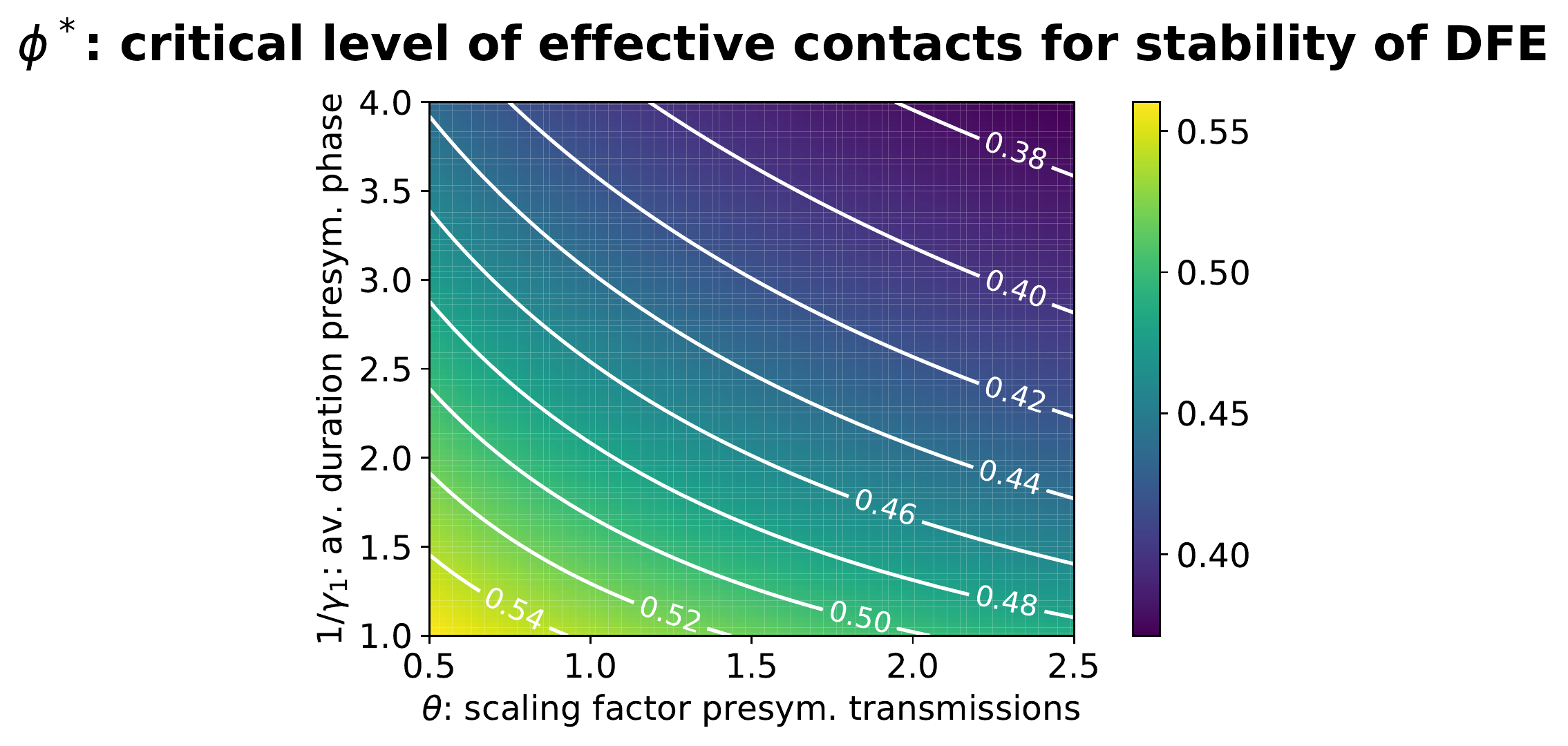}
		\caption{{\bf Critical level $\phi^*$ of effective contacts for the stability of the DFE varying disease characteristics.} Here we vary the average duration of the early infectious period $1/\gamma_1$ and the scaling factor $\theta$ for the early infectious transmission rate.}
		\label{fig:crit_red_dis_char1}
		\centering
	\end{figure}
	
	The extent of transmission during the early infectious phase is characterized by the average length of the early infectious period $1/\gamma_1$, and by the scaling factor $\theta$ describing the relative transmission rate of early infectious individuals when compared to late infectious individuals. We investigated the effects of variations in these parameters on  $\phi^*$. When varying $1/\gamma_1$, we adjusted $1/\gamma_2$ such that we maintain the same average for the total duration of the infectious period $1/\gamma=1/\gamma_1+1/\gamma_2$, and adjusted the transmission rates in order not to alter $\mathcal{R}_0$. Both larger $1/\gamma_1$ and larger $\theta$ increase the amount of transmissions that cannot be prevented by symptom-based testing. Larger $1/\gamma_1$ additionally implies a smaller detection ratio achieved by testing (at least if $1/\gamma_1+1/\gamma_2$ and $\sigma_{U_2}$ stay unchanged) and a larger testing delay such that contacts have spent more time in the chain of infection before being quarantined. As shown in \autoref{fig:crit_red_dis_char1}, variations in both parameters notably affect $\phi^*$.
	\begin{figure}
		\centering
        \includegraphics[width=\linewidth]{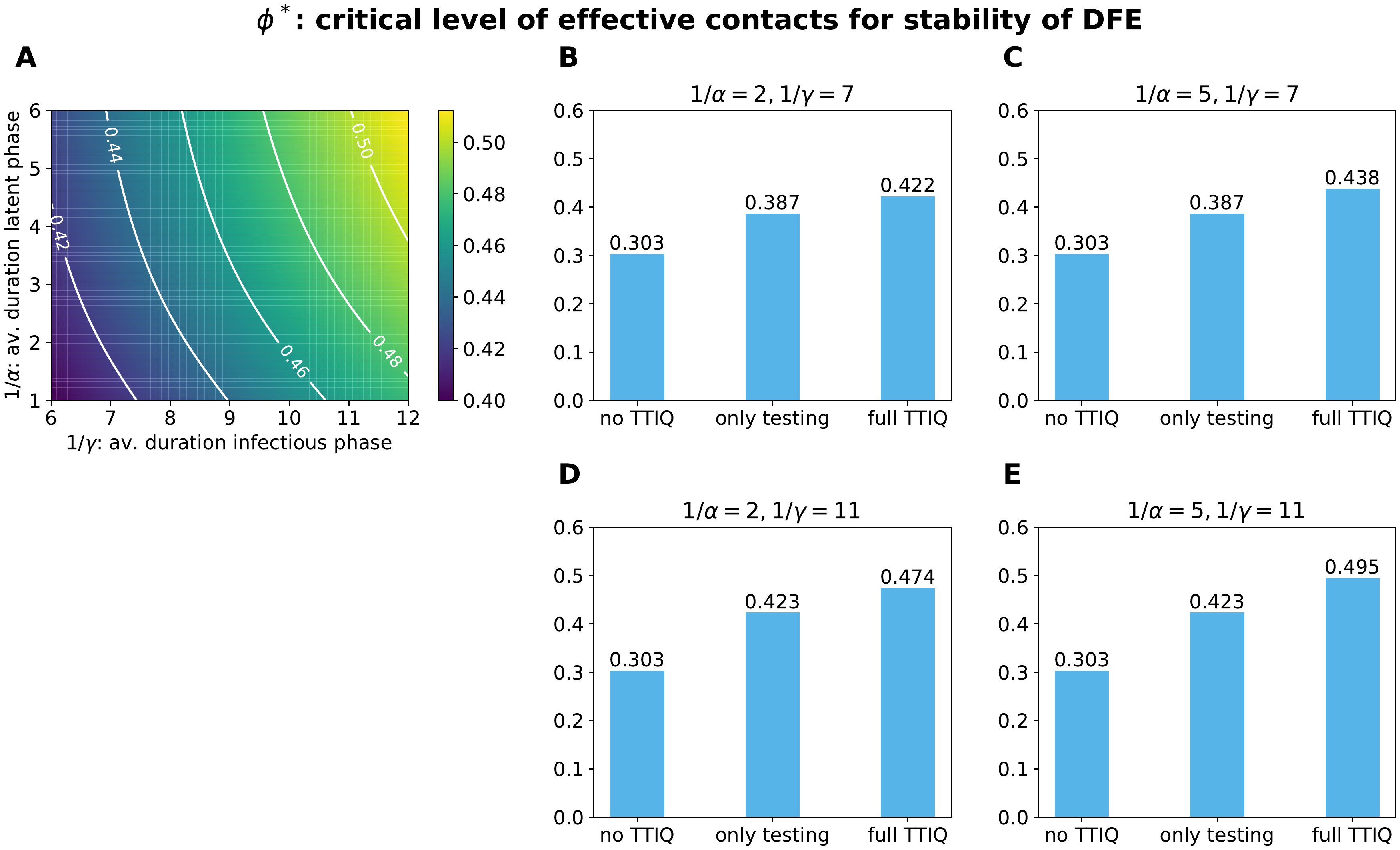}
		\caption{{\bf Critical level $\phi^*$ of effective contacts for the stability of the DFE varying disease characteristics.} Here we vary the average duration of the latent phase $1/\alpha$ and the average duration of the infectious period $1/\gamma$. This is shown for \textbf{A} the $(1/\gamma,1/\alpha)$-plane and \textbf{B}-\textbf{E} different combinations of values for $1/\alpha$ and $1/\gamma$, comparing a scenario without any TTIQ to a scenario where only testing is conducted and a scenario where both testing and contact tracing is carried out.}
		\label{fig:crit_red_dis_char2}
		\centering
	\end{figure}
	
    To assess the sensitivity of TTIQ to the time scale of disease spread, we also considered changes in the mean duration of the latent period $1/\alpha$ and the mean duration of infectious phase $1/\gamma=1/\gamma_1+1/\gamma_2$. When varying $1/\gamma$ we adjusted the transmission rate to maintain $\mathcal{R}_0$ constant and kept a constant ratio between $\gamma_1$ and $\gamma_2$ such that the proportion of transmissions occurring in the early infectious phase in the absence of TTIQ remains unchanged. Unsurprisingly, faster disease progression (smaller $1/\alpha$, $1/\gamma$) leads to a smaller $\phi^*$ (\autoref{fig:crit_red_dis_char2}). A longer latency period (larger $1/\alpha$) implies that more infected contacts are still latent at the time of being traced and thus more onward transmissions are prevented by their quarantine. This only affects the effectiveness of contact tracing and has no effect on a purely testing-based scenario (compare second bars in \autoref{fig:crit_red_dis_char2}B-E). Larger $1/\gamma$ increases the detection ratio achieved by testing (at least if $\sigma_{U_2}$ stays unchanged) and implies that less contacts will have recovered by the time of being quarantined. Figure \ref{fig:crit_red_dis_char2}B-E demonstrates that this increases the benefit of both testing and contact tracing. 
	\begin{figure}
		\centering
        \includegraphics[width=0.75\linewidth]{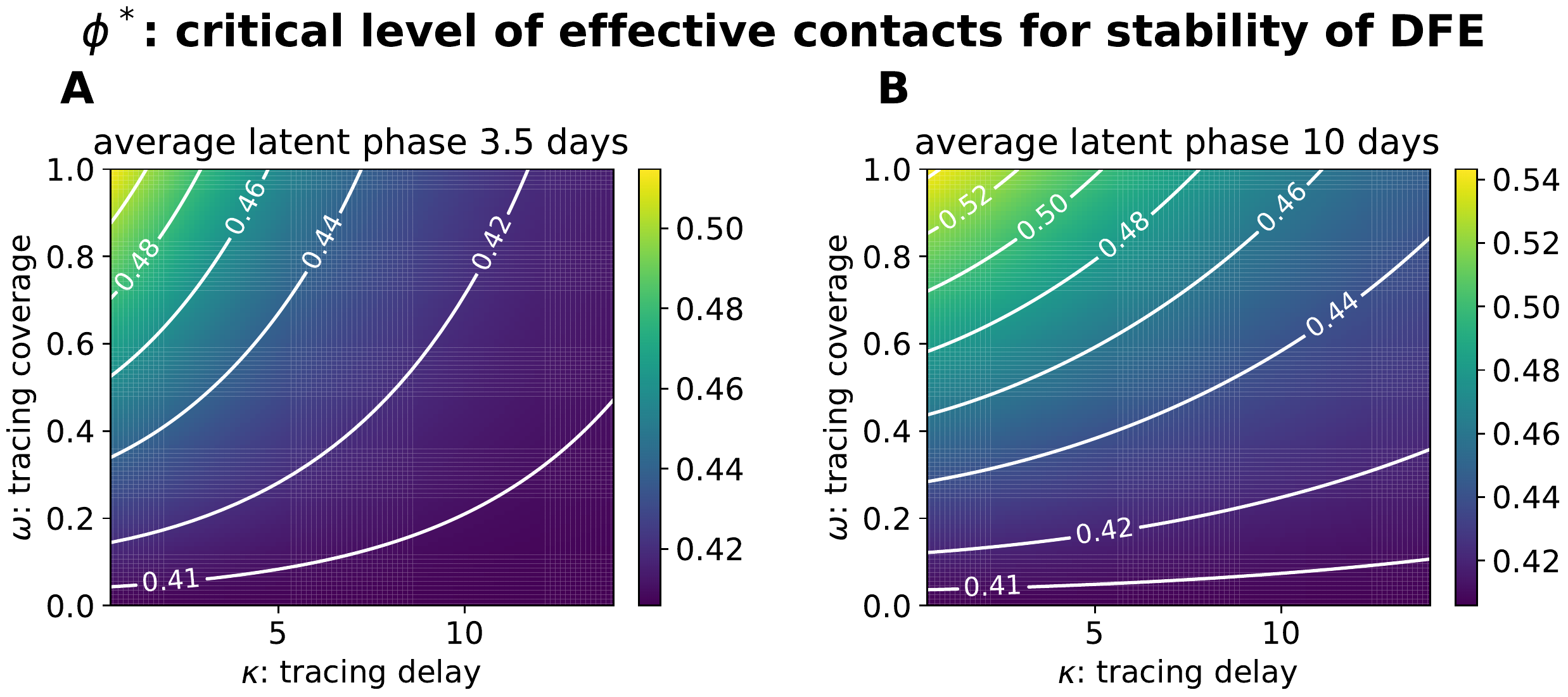}
		\caption{{\bf Critical level $\phi^*$ of effective contacts for the stability of the DFE varying the tracing delay $\kappa$ and the tracing coverage $\omega$.} This is shown assuming an average latent phase of either \textbf{A} $3.5$ days or \textbf{B} $10$ days.}
		\label{fig:crit_red_dis_char3}
		\centering
	\end{figure}
	
	We found that disease characteristics not only influence $\phi^*$ directly but also its sensitivity with respect to the TTIQ parameters. For instance, the tracing delay (at least for the values considered here) gains relevance when compared to the tracing coverage in case of a shorter latent phase (compare slope of contour lines in \autoref{fig:crit_red_dis_char3}A and \autoref{fig:crit_red_dis_char3}B). 
	\subsubsection*{Diminishing effectiveness of TTIQ during outbreak}
	When the disease prevalence rises the effectiveness of testing and tracing decreases and our stability analysis about the DFE does not offer sufficient information for disease control. To investigate how the effectiveness of TTIQ is affected by increasing prevalence, we considered a scenario where we assume moderately effective social and hygiene measures that are insufficient to control disease spread at the DFE ($\phi=0.6>0.46\approx\phi^*$). We initiated a simulation of our model near the DFE by choosing a constant history function such that the simulation starts of with an incidence of roughly $300$ confirmed cases per day. Although not being an exact representation, this setting is qualitatively resembling the surge in cases in late summer and early fall of 2020 in Germany: most of the population is still susceptible, the simulation starts at a low daily incidence of confirmed cases and disease spread is moderately suppressed by contact restrictions. For simplicity, and recognizing that transmission rates and TTIQ effort change much more dynamically in reality, all parameters in this scenario are held constant throughout the simulated period with values as in the baseline setting \autoref{tab:para}. 
    \begin{figure}
		\centering
		\includegraphics[width=0.75\linewidth]{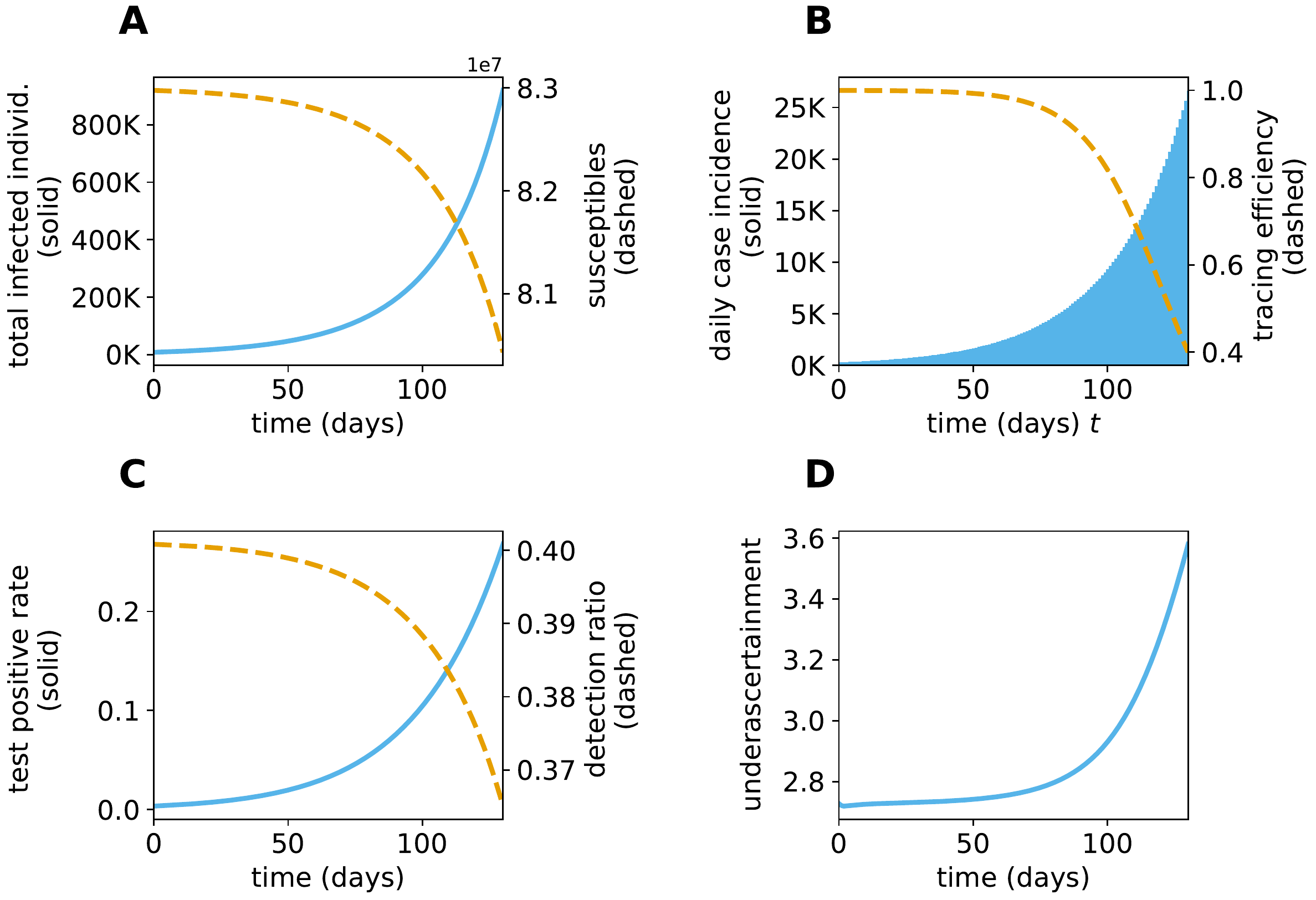}
		\caption{{\bf Model outcomes simulating an outbreak starting from low case numbers.} This simulation is initiated using the baseline parameter values given in \autoref{tab:para} and a reduction of effective contacts corresponding to $\phi=0.6$.}
		\label{fig:rising}
	\end{figure}
	In the considered scenario the DFE is unstable leading to an increase in the total number of infected individuals (\autoref{fig:rising}A) and daily new confirmed cases (\autoref{fig:rising}B) over time. As more infectious individuals arise in the population the test positive rate increases (\autoref{fig:rising}C). Moreover, the finite testing capacity leads to decreasing per capita detection rates which decreases the detection ratio $$\frac{\eta_{U_1}(t)}{\gamma_1+\eta_{U_1}(t)} + \frac{\gamma_1}{\gamma_1+\eta_{U_1}(t)} \frac{\eta_{U_2}(t)}{\gamma_2+\eta_{U_2}(t)}.$$ The nevertheless increasing number of confirmed cases leads to an increase in reported contacts which gradually diminishes the tracing efficiency (\autoref{fig:rising}B)  $$\varepsilon(t)=\frac{\Omega}{(c_\text{pot}(t)^2 + \Omega^2)^{1/2}}.$$ All in all these effects lead to a steep increase in the under-ascertainment of active infectious individuals $$\frac{U_1(t)+U_2(t)+Q_{U_1}(t)+Q_{U_2}(t)+I_1(t)+I_2(t)}{Q_{U_1}(t)+Q_{U_2}(t)+I_1(t)+I_2(t)}$$ as prevalence rises (\autoref{fig:rising}D). This leads to a self-accelerating disease spread that gets harder to control the longer it remains uncontrolled. This is reflected in the critical level $\overline{\phi}$ of effective contacts that yields a timely stagnation in the incidence of infected individuals. We calculated $\overline{\phi}$ at multiple points in time $t^*$ by simulating an intervention that immediately decreases $\phi$, that is
	$$\phi(t)=\begin{cases}
		0.6,\quad t<t^*\\
		\phi_1,\quad t\geq t^*
	\end{cases}
	$$
	and scanned for the maximal value of $\phi_1$ that yields no further increase (thus approximately yields stagnation) in the incidence after two weeks post intervention for the rest of the additionally simulated period ($+42$ days). The result is shown by the dashed line in \autoref{fig:crit_red_timing}A.
	\begin{figure}
		\centering
		\includegraphics[width=0.65\linewidth]{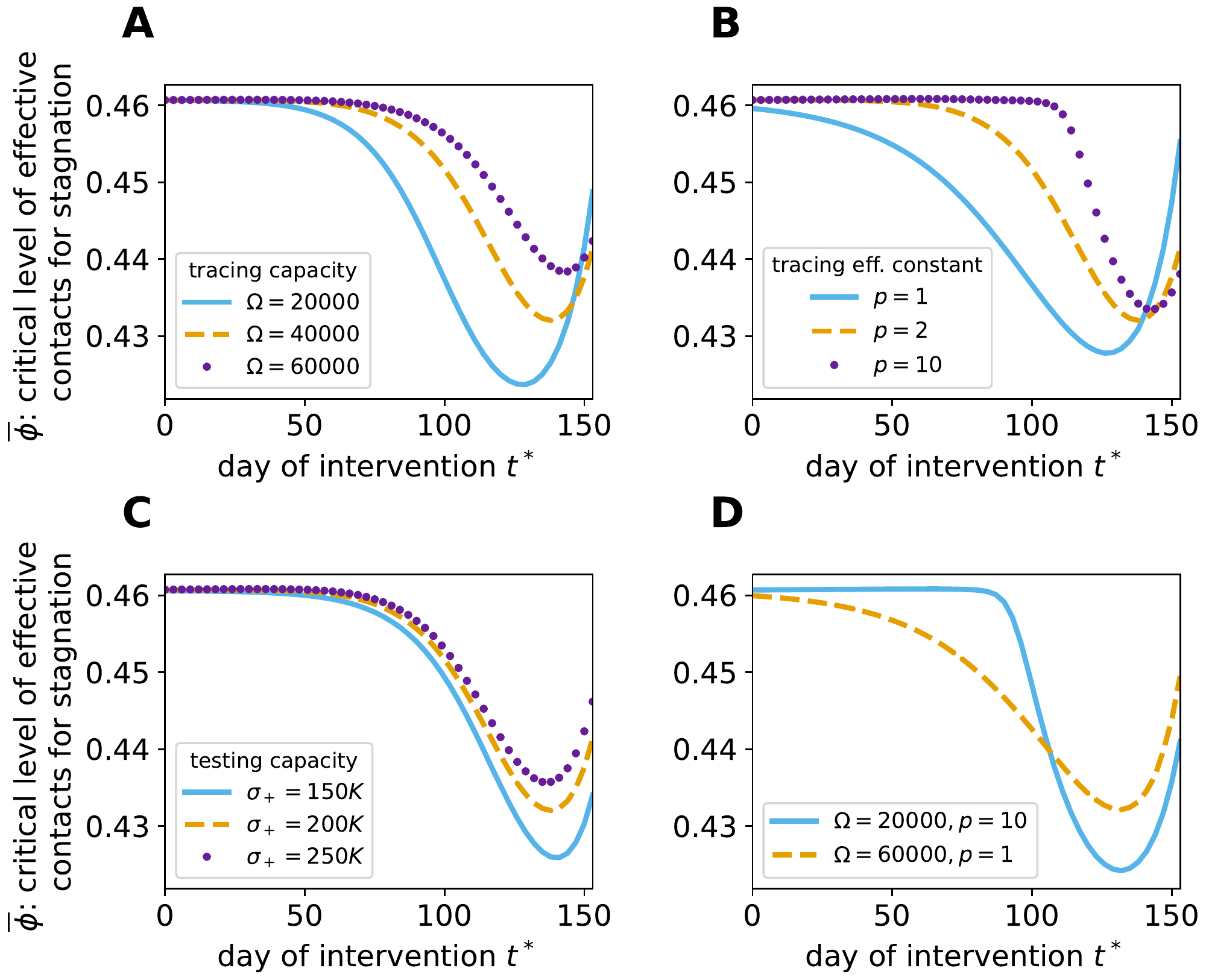}
		\caption{{\bf Critical level $\overline{\phi}$ of effective contacts that yields a stagnation in the incidence of infected individuals.} $\overline{\phi}$ is calculated along the epidemic outbreak considered in \autoref{fig:rising} and plotted against the timing of intervention $t^*$. This is shown for different values of \textbf{A} the tracing capacity $\Omega$, \textbf{B} the tracing efficiency constant $p$, \textbf{C} the testing capacity $\sigma_+$ and \textbf{D} comparing a scenario with high tracing capacity $\Omega$ and low tracing efficiency constant $p$ to a scenario with low $\Omega$ and large $p$.}
		\label{fig:crit_red_timing}
	\end{figure}
	Due to the loss of TTIQ effectiveness, later intervention timing $t^*$ requires a stricter reduction of effective contacts (lower $\overline{\phi}$) in order to prevent a further increase in the incidence of infected individuals. However, at some point the effect of population-level immunization (or depletion of susceptibles) counterbalances the effect of a decreasing TTIQ effectiveness and $\overline{\phi}$ starts increasing as $t^*$ becomes larger (this ease of control comes at the cost of widespread infestation). Models that do not include limited capacities would predict this increase in $\overline{\phi}$ already for early intervention time points $t^*$.
	\subsubsection*{Impact of capacity parameters}
	We investigated the impact of TTIQ capacities on the temporal evolution of $\overline{\phi}$ by simulating the epidemic outbreak considered in \autoref{fig:rising} for multiple values of the tracing capacity $\Omega$, the tracing efficiency constant $p$, and on the testing capacity $\sigma_+$. All three, larger tracing capacity $\Omega$, larger testing capacity $\sigma_+$, and larger tracing efficiency constant $p$ (reflecting higher efficiency in allocating tracing capacity) increase the minimal value of $\overline{\phi}$ along the outbreak and delay its decrease with respect to $t^*$ (\autoref{fig:crit_red_timing}A-C). However, for larger $p$ the decrease in $\overline{\phi}$ happens more abruptly. In particular, a TTIQ system with relatively low tracing capacity but with efficient allocation of this capacity (large $p$) manages to maintain a high tracing efficiency for a longer phase of the outbreak than a TTIQ system with high tracing capacity but inefficient allocation of this capacity (small $p$). This is only true until a certain prevalence is reached. The system with low absolute tracing capacity experiences the drop in $\overline{\phi}$ in a shorter time frame and quite abruptly stricter reductions of effective contacts are needed to regain control over disease spread (\autoref{fig:crit_red_timing}D). 
 	\subsubsection*{Implications for intervention strategies}
\begin{figure}
 		\centering
 		\includegraphics[width=\linewidth]{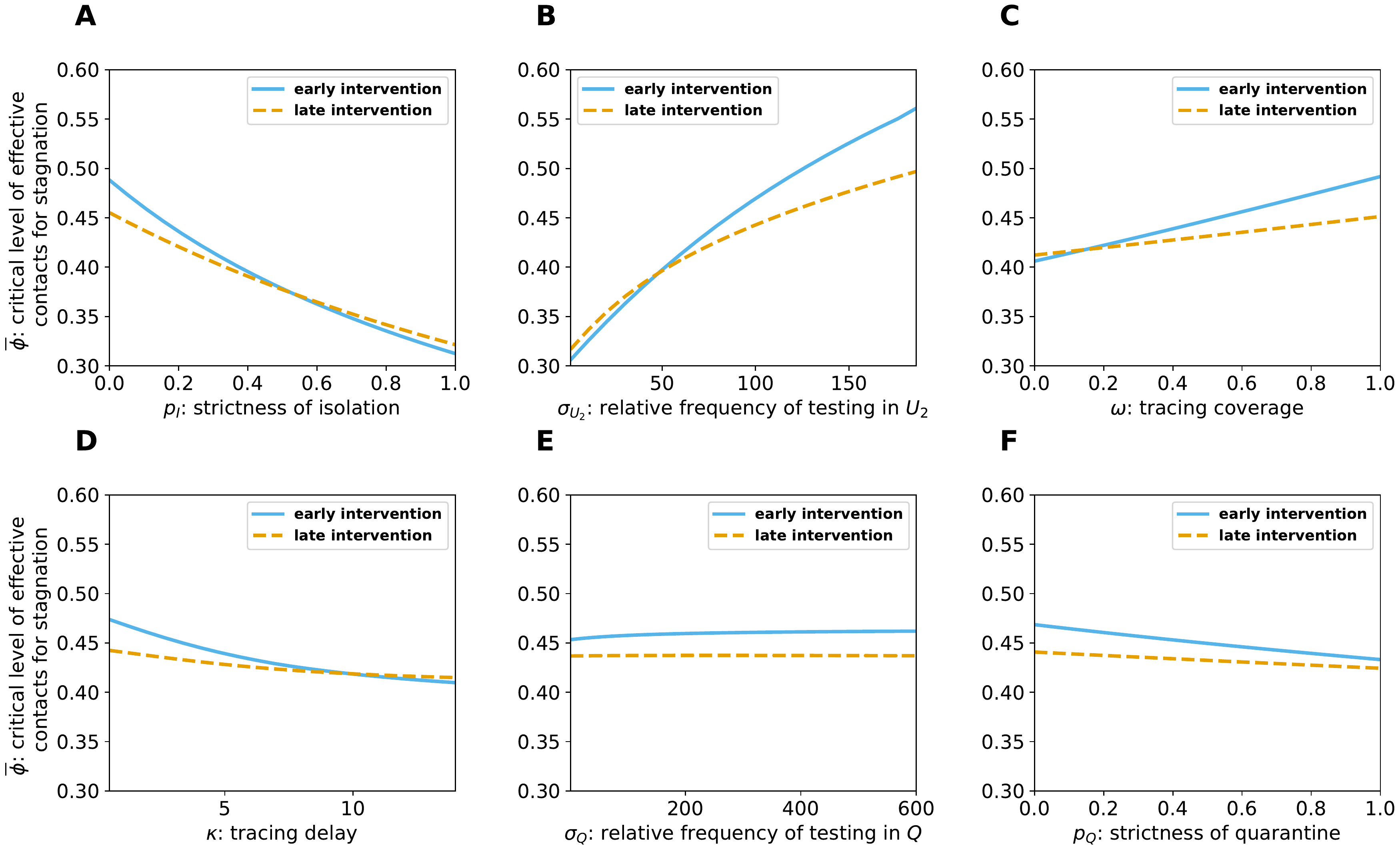}
 		\caption{{\bf Critical level $\overline{\phi}$ of effective contacts that yields a stagnation in the incidence of infected individuals varying single TTIQ parameters.} This is shown for an early intervention time point corresponding to a daily case incidence of approximately  $1\,500$ (solid lines) and for a late intervention time point corresponding to a daily case incidence of approximately $20\,000$ (dashed lines).}
 		\label{fig:crit_red_pseudo}
 	\end{figure}

    The decrease in TTIQ effectiveness along an epidemic wave implies that the success of intervention strategies is state-dependent. To see how this affects interventions based on changes in TTIQ parameters we considered the outbreak investigated before (see \autoref{fig:rising}) and simulated an intervention taking place either early at a daily case incidence of approximately $1\,500$ ($t^*=47$) or late at at a daily case incidence of approximately  $20\,000$ ($t^*=123$). At the start of the intervention we varied TTIQ parameters in ranges as in \autoref{tab:param_ranges} and additionally scanned for the associated critical level $\overline{\phi}$ of effective contacts that yields a stagnation in the incidence of infected individuals (\autoref{fig:crit_red_pseudo}).
    While the curves corresponding to the early intervention unsurprisingly resemble those in \autoref{fig:crit_red_DFE}, the curves corresponding to the late intervention scenario are shifted. For optimistic TTIQ parameter values, the curves are shifted downwards, reflecting the decreased tracing efficiency and per capita testing rates. Conversely, for suboptimal TTIQ parameter sets, $\overline{\phi}$ is larger in the late intervention scenario due to the already greater depletion of susceptibles. In these cases, the beneficial effect of higher immunization outweighs the detrimental effect of lower TTIQ effectiveness, since TTIQ does not contribute much in the first place. For all TTIQ parameters, the curve corresponding to the late intervention is notably flatter, demonstrating less impact on $\overline{\phi}$ and decreased importance of these parameters for disease control. 
 	 
\begin{figure}
 		\centering
		\includegraphics[width=0.75\linewidth]{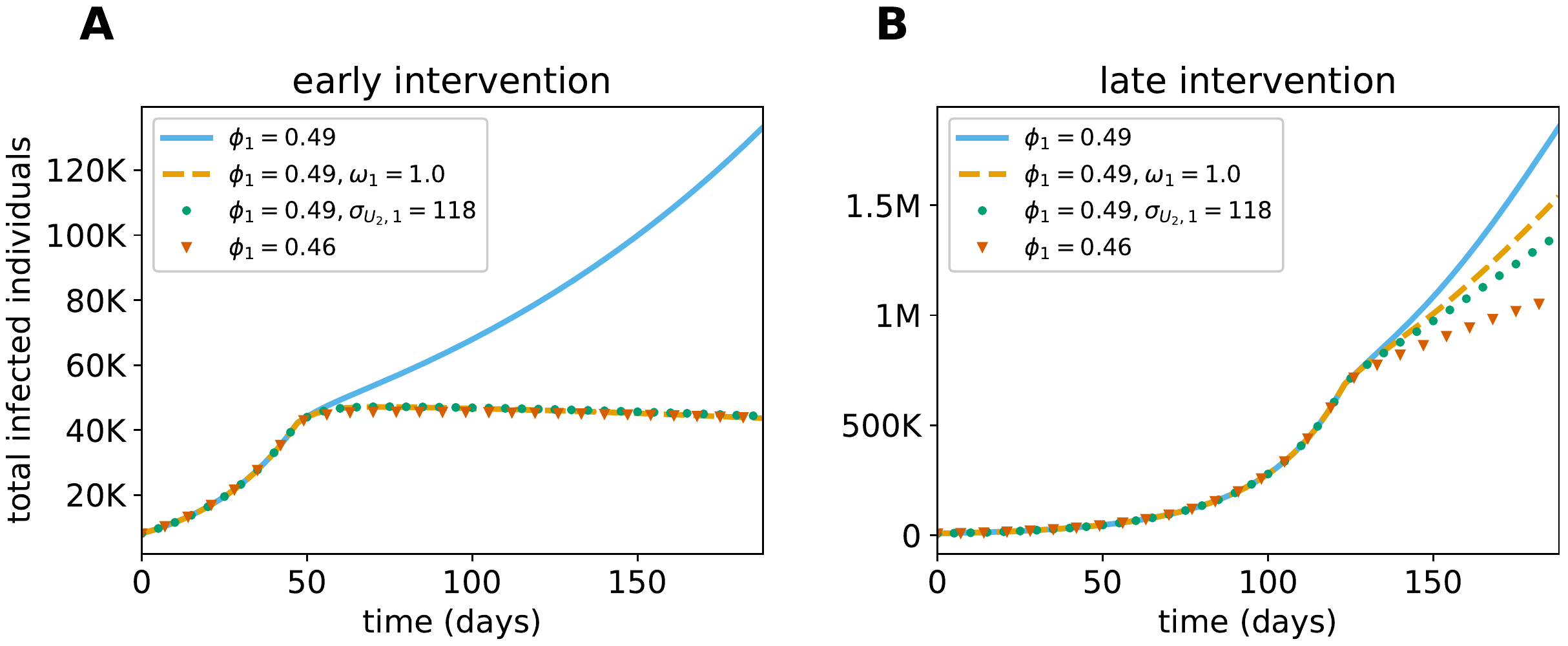}
 		\caption{{\bf Number of infected individuals (both detected and undetected) simulating the baseline outbreak scenario from \autoref{fig:rising} with different interventions.} Here an intervention takes place either \textbf{A} early at a daily case incidence of approximately $1\,500$ ($t^*=47$) or \textbf{B} late at a daily case incidence of approximately $20\,000$ ($t^*=123$). At the time of intervention the level of effective contacts is changed from its baseline value to $\phi_1$. Additionally, in some considered scenarios, the tracing coverage is improved to $\omega_1$ or the relative frequency of testing undetected infectious individuals to $\sigma_{U_2,1}$. Parameters not mentioned in the legend are held constant throughout the simulation.}
 		\label{fig:strategies}
 	\end{figure}

    The above observations have important implications for the outcome of intervention strategies. To demonstrate that, we report here simulated scenarios where at the day of intervention a stricter reduction of effective contacts (smaller $\phi$) is applied and, additionally, for some considered scenarios different TTIQ parameters are improved. As before we assumed that the interventions have an immediate effect on the parameters of the system. As a benchmark scenario we assumed that the intervention (applied either early or late as above) leads to a reduction in effective contacts from $\phi=0.6$ to $\phi_1=0.49$. When accompanied with no additional improvements of TTIQ, disease spread is controlled in neither the early nor the late intervention scenario (see solid lines in \autoref{fig:strategies}).
    Additionally increasing the tracing coverage from $\omega=0.65$ to $\omega_1=1$ (e.g., by increasing awareness in the population to keep track of personal contacts or by improving the close contact definition), improving the relative frequency of testing undetected late individuals from $\sigma_{U_2}=93$ to $\sigma_{U_2,1}=118$, or implementing a stricter reduction of effective contacts corresponding to $\phi_1=0.46$, show a similar response in case of low prevalence and timely lead to a slow decrease in the number of infected individuals (\autoref{fig:strategies}A). However, the three enhanced interventions show different responses in the late intervention scenario with high prevalence. Improving the tracing coverage proves to be ineffective (see dashed line \autoref{fig:strategies}B). The improvement in testing does not perform significantly better (see dotted line in \autoref{fig:strategies}B). Only the stricter reduction of effective contacts significantly slows down disease spread (see triangles in \autoref{fig:strategies}B). All three strategies, however, fail to stop the increase in the number of infected individuals in the late intervention scenario.
	
\section{Discussion}
    We have introduced a delay differential equation model to assess the effectiveness of TTIQ interventions for infectious disease control. To account for limited testing capacity, we introduced state-dependent detection rates of infectious cases that are based on the derivation presented in \cite{Barbarossa2021}. This leads to reasonably high detection rates at low prevalence that smoothly decrease when prevalence rises. Similar to the approach in \cite{Contreras2021a}, we model contact tracing as a delayed consequence of successful index case identification. However, in our derivation we did not consider a sharp prevalence threshold above which the tracing efficiency is affected by disease spread. Instead, by contrasting the theoretical yield of detected infections per index case and the curtailment of this yield by the emerging burden on PHA, our model rather describes a smoothly decreasing tracing efficiency as disease prevalence rises. In addition, our model includes an early infectious phase in the course of infection during which infected individuals can transmit the disease prior to the occurrence of potential symptoms.
    
    As a working example representative of a (re-)emerging disease for which pharmaceutical interventions are not yet available, we applied our model to study the effectiveness of TTIQ in the context of the spread of COVID-19 in Germany during the wave in late summer and fall of 2020. Under these conditions, in particular original strain of SARS-CoV-2, no vaccine, no rapid antigen tests, and low population immunity, our model results suggest that, as long as disease prevalence is low, TTIQ allows to control disease spread if the reproduction number is reduced to a value below $1.52$ by other interventions alone. Thus, assuming a basic reproduction number of $3.3$, TTIQ as described here would allow for approximately $52\%$ more effective contacts within the population. Nevertheless, a reduction to approximately $46\%$ of the pre-COVID-19 effective contact rate would be needed to prevent an epidemic outbreak. This is in line with previous modeling studies suggesting that  additional interventions are required to control disease spread under realistic assumptions on imperfect TTIQ \cite{Ferretti2020, Kerr2021, Pollmann2021, Davis2021, Hellewell2020, Kretzschmar2021, Kretzschmar2020, Kucharski2020, Contreras2021a, Contreras2021, Scarabel2021}. 
    
    By means of a sensitivity analysis we identified the TTIQ parameters most influential to the effectiveness of TTIQ. Our results show that depending on how the TTIQ parameters are set, TTIQ may allow a significantly higher or lower effective contact rate than observed in our baseline setting. In agreement with previous findings in the literature, the compliance with isolation \cite{Kerr2021, Quilty2021, Davis2021} and the rate of index case identification  \cite{Kerr2021, Pollmann2021, Ashcroft2022, Davis2021, Hellewell2020, Kretzschmar2020, Browne2015} play a central role in this regard. The significance of these parameters reflects simple causal relationships between the mechanisms described by the TTIQ parameters. It is irrelevant how many index cases are found by testing and how many contacts are traced when none of them effectively reduces their contacts. Similarly, by definition contacts can only be traced and quarantined upon prior identification of index cases. This renders the success of testing a key factor for the effectiveness of TTIQ. Consistent with previous studies \cite{Browne2015, Contreras2021, Sturniolo2021}, we observed synergistic effects when varying TTIQ parameters simultaneously which highlights the benefit of combining strategies rather than concentrating on, e.g., testing exclusively. 
    
    To go beyond our consideration of COVID-19 and to further examine the effect of uncertainty in our baseline parameter setting, we also considered variations in parameters describing disease characteristics. Overall, our results underline that COVID-19 combines several characteristics adverse to TTIQ (asymptomatic and early transmissions, short latency and infectious period, airborne transmission resulting in imperfect tracing coverage) that explain its rather low effectiveness observed in our baseline setting. The central role of asymptomatic and presymptomatic transmission in diminishing TTIQ effectiveness has been repeatedly demonstrated in the literature \cite{Jarvis2021, Pollmann2021, Hellewell2020, Kretzschmar2021, Kucharski2020, Browne2015, Fraser2004}.
    
    When disease control is insufficient and an outbreak takes place, limited TTIQ capacities lead to a self-acceleration of disease spread \cite{Contreras2021a, Contreras2021, Lunz2021, Scarabel2021}. In our model, this effect takes place gradually along an epidemic wave until it is countered by sufficiently wide spread immunization. We show how the timing and intensity of the self-accelerating effect depend on the capacity parameters. It is weaker for higher capacities and a more efficient allocation of contact tracing expressed by larger tracing efficiency constant $p$. Moreover, a larger $p$ extends the period of almost perfect tracing efficiency and leads to a more abrupt decrease of tracing efficiency. The limit $p\to\infty$ offers a transition between the description of tracing capacity in our model and previous models that assume a sharp prevalence threshold for the decrease in tracing efficiency \cite{Contreras2021a, Contreras2021, Scarabel2021}. 
    
    The self-accelerating effect observed in our working example appears to be only moderate. Our sensitivity results and the previous literature shows that it may be more or less pronounced depending on assumptions on the model structure and parameters (see for example \cite{Contreras2021a}). It should be noticed, however, that the detrimental effect of limited capacity in our simulations, though less pronounced than under alternative modeling assumptions, is far from negligible and the transient acceleration of disease spread may have irreversible effects \cite{Scarabel2021}. In particular, our results show that at states of high prevalence stricter measures are needed to control disease spread and interventions based on improvements of TTIQ parameters become less effective. In such situations, greater reduction of transmission rates by means of stricter social or hygiene measures might be the only feasible non-pharmaceutical intervention to effectively stop case numbers from rising. 
   
    There are several limitations to our model that offer routes for further research and should be considered when interpreting our results. We focused on first-order manual forward tracing. Other modeling studies consider the effect of recursive tracing \cite{Kiss2007, Pollmann2021, Klinkenberg2006, Mueller2016}, backward tracing \cite{Kojaku2021, Mueller2016, Mueller2000} and digital contact tracing \cite{Ferretti2020, Pollmann2021, Kretzschmar2020, Kucharski2020}. Under favorable conditions, like a high acceptance within the population in case of digital contact tracing, these processes may significantly increase the effectiveness of TTIQ. Moreover, in reality, a relevant proportion of contacts of confirmed cases is likely to be informed about their potential transmission by the index case before PHA reach out to the contact. This can lead to earlier quarantine of contacts and render parts of contact tracing independent of PHA capacities. Similarly, individuals might reduce their contacts due to an increase in reported cases. These and other behavioral factors are so far not considered in our model. Moreover, we only considered quarantine for close contacts that happened to be infected by one of the identified index cases. It should be noticed that quarantine of the remaining contacts (those that had contact but were not infected), which is for example considered in \cite{Pollmann2021, Lunz2021, Sturniolo2021, Webb2015}, can effect the disease dynamics if sufficiently many susceptibles are quarantined or a substantial amount of additional infected individuals are quarantined by chance (this becomes relevant only at a high prevalence). In addition, the consideration of quarantine of all close contacts (infected and uninfected) would allow to assess the socioeconomic damage induced by contact tracing and to address TTIQ strategies that reduce quarantine costs (see for example \cite{Quilty2021, Kucharski2020, Lunz2021, Ashcroft2021}). Furthermore, our approach considers the populations in the different compartments in our model as homogeneous. At the cost of an extended parameter space many additional factors that differentiate individuals could be considered, such as age, space, overdispersion and disease severity. Not differentiating infectious individuals by the severity of their disease, for instance, implies that the significant benefit of increasing the relative frequency $\sigma_{U_2}$ of testing undetected late infectious individuals when compared to susceptibles shown in \autoref{fig:crit_red_DFE}C should be seen as harder to realize the larger $\sigma_{U_2}$ becomes. Every increase of $\sigma_{U_2}$ deviates the typical distribution of individuals in $U_2$ more and more towards asymptomatic individuals. This makes the average individual in $U_2$ less amenable to symptom-based testing and a further increase of $\sigma_{U_2}$ more difficult to achieve. Furthermore, our approach assumes homogeneous mixing. However, the effect of contact tracing is dependent on the contact network underlying the considered population. Previous modeling studies found that, for instance, clustering benefits contact tracing \cite{Eames2003, House2010} and that mixing patterns can affect the efficacy of contact tracing \cite{Kiss2007}. Additionally, although our model splits the infectious phase into an early and late infectious stage, a more realistic infectivity profile and course of symptoms could be considered using an age of infection approach \cite{Ferretti2020, Pollmann2021, Mueller2016, Mueller2000, Fraser2004, Huo2015, Scarabel2021}. If the state-dependent dynamics induced by limited capacities could be incorporated as detailed as in our model, an age of infection approach would offer a natural environment to overcome many approximations that we applied in the derivation of the contact tracing terms. An agent-based framework \cite{Jarvis2021, Kerr2021, Pollmann2021, Quilty2021} would allow to include even more complexity and to consider various of the factors outlined above. A comparison to such complex formulations of contact tracing could be used to investigate the justification of a numerically inexpensive but approximate model as leveraged in the present work.
        
\section{Conclusion}
    We rigorosuly derived a detailed mechanistic model of TTIQ interventions that accounts for challenges posed by disease characteristics and inherent limitations such as a tracing delay, imperfect compliance with isolation, and limited testing and tracing resources. Using the spread of COVID-19 as an example, we show how these factors can limit the effectiveness of TTIQ as a strategy for disease control. Our observations on the diminishing TTIQ effectiveness during simulations of an epidemic outbreak, demonstrate that a careful evaluation of the contemporary load on TTIQ capacities is needed to predict the effect of different intervention strategies. A strength of our approach is that we disentangle the individual contributions to disease control that result from isolating index cases and tracing their contacts. Our model extends the state-of-the-art in mean field models of TTIQ and is flexible enough to be adapted and applied to evaluate the effectiveness of TTIQ in controlling infectious diseases other than COVID-19.
	
	\section*{Acknowledgments} 
	MVB and JH were supported by the LOEWE focus CMMS.
	
	\section*{Conflict of interest}
	We have no conflict of interest to declare.
	
	\section*{Code availability}
	The code necessary to reproduce the results presented in this work is available on GitHub \small \url{https://github.com/julehe/TTIQ}\normalsize.
	
	\providecommand{\href}[2]{#2}
	\providecommand{\arxiv}[1]{\href{http://arxiv.org/abs/#1}{arXiv:#1}}
	\providecommand{\url}[1]{\texttt{#1}}
	\providecommand{\urlprefix}{URL }

\begin{appendices}

\section{Derivation of contact tracing terms with early and late infectious individuals}\label{app:presym}
	In this section we outline the derivation of the terms $Tr_E,Tr_{U_1},Tr_{U_2}$ describing contact tracing in the full model \eqref{eq:fullsys_presym}. To this end, we distinguish between index cases detected while being in the late infectious phase $U_1$ and those being detected while being in the late infectious phase $U_2$. On average, these different types of index cases report different numbers of infected contacts, and their contacts spent a different duration in the infected chain at the time of being quarantined.
		    
	First, consider an average index case detected by testing at time $t-\kappa$ while being in $U_2$ (here called {\it $U_2$-index case}). Following the same reasoning as in Section \ref{sec:tracing}, we assume that the PHA choose a tracing window $T$ and ask for close contacts from the tracing interval $J^{U_2}_{t,T}=[t-\kappa-T,t-\kappa]$. The duration $\tau^{U_2}_t$ for which the index case has been infectious by the time of detection $t-\kappa$ is the sum of the time spent in $U_1$ and the time spent in $U_2$, $$\tau^{U_2}_t=\tau^{U_2}_{t,U_1}+\tau^{U_2}_{t,U_2}.$$ Depending on the tracing window $T$, the tracing interval $J^{U_2}_{t,T}$ is composed of three subintervals: 
	\begin{itemize}
	    \item the (potentially trivial) part of $J^{U_2}_{t,T}$ during which the index case was not yet infectious
	    $$J^{U_2,\text{ninf}}_{t,T}=\begin{cases}
		\emptyset, &\quad \text{if}\,\, T\leq\tau^{U_2}_t\\
		[t-\kappa-T,t-\kappa-\tau^{U_2}_t], &\quad \text{if}\,\, T>\tau^{U_2}_t
	    \end{cases}$$
	    
	    \item the (potentially trivial) part of $J^{U_2}_{t,T}$ where the index case was in the early infectious phase $U_1$
	    \begin{align}\label{eq:interv_U2pre}
		J^{U_2,\text{early}}_{t,T}=\begin{cases}
			\emptyset, &\quad \text{if}\,\, T\leq\tau^{U_2}_{t,U_2}\\
			[t-\kappa-T,t-\kappa-\tau^{U_2}_{t,U_2}],&\quad \text{if}\,\, \tau^{U_2}_{t,U_2}<T\leq\tau^{U_2}_t\\
			[t-\kappa-\tau_t^{U_2},t-\kappa-\tau^{U_2}_{t,U_2}], &\quad \text{if}\,\, T>\tau^{U_2}_t
		\end{cases}
	    \end{align}
	    
	    \item the time in $J^{U_2}_{t,T}$ during which the index case was in the late infectious phase $U_2$
	    \begin{align}\label{eq:interv_U2pos}
		J^{U_2,\text{late}}_{t,T}=\begin{cases}
			[t-\kappa-T,t-\kappa], &\quad \text{if}\,\, T\leq\tau^{U_2}_{t,U_2}\\
			[t-\kappa-\tau^{U_2}_{t,U_2},t-\kappa],&\quad \text{if}\,\, T>\tau^{U_2}_{t,U_2}.
		\end{cases}
	    \end{align}
	\end{itemize}
	In reality, the close contact definition probably leads to different reported close contact rates $\widetilde{c}_{0},\widetilde{c}_{1},\widetilde{c}_{2}$ corresponding to $J^{U_2,\text{ninf}}_{t,T}, J^{U_2,\text{early}}_{t,T}, J^{U_2,\text{late}}_{t,T}$ and different infection probabilities $\widetilde{p}_{1},\widetilde{p}_{2}$ corresponding to $J^{U_2,\text{early}}_{t,T}, J^{U_2,\text{late}}_{t,T}$. These result in transmission rates observed by contact tracing $\widetilde{\beta}_{1}:=\widetilde{p}_{1}\,\widetilde{c}_{1}$ in $J^{U_2,\text{early}}_{t,T}$ and $\widetilde{\beta}_{2}:=\widetilde{p}_{2}\,\widetilde{c}_{2}$ in $J^{U_2,\text{late}}_{t,T}$. Following our argumentation in Section 2.2, we make the simplifying assumption that $\widetilde{c}:=\widetilde{c}_{0}=\widetilde{c}_{1}=\widetilde{c}_{2}$. Additionally, we set $\widetilde{p}_{1}=\theta\widetilde{p}_{2}$, where $\theta$ is the scaling factor for the early infectious transmission rate. Note that this results in
	\begin{align}\label{eq:assum_betas}
		\frac{\widetilde{\beta}_{1}}{\beta_{U_1}}=\frac{\widetilde{\beta}_{2}}{\beta_{U_2}}.
	\end{align} 
	Following the approach in Section \ref{sec:tracing}, we approximate the rate at which contacts of $U_2$-index cases become traceable at time $t$ as
	\begin{align}\label{eq:U2_contacts}
		c^{U_2}_{\text{pot}}(t)=\abs{J^{U_2}_{t,T}}\,\widetilde{c}\,\eta_{U_2}(t-\kappa)U_2(t-\kappa).
	\end{align}
	The rate at which infected contacts of $U_2$-index cases become traceable at time $t$ is the sum of those infected while their $U_2$-index case was in $U_1$ 
	\begin{align*}
		c^{\text{inf},U_2}_{\text{pot},U_1}(t)=\frac{\abs{J^{U_2,\text{early}}_{t,T}}}{\abs{J^{U_2}_{t,T}}}\,\widetilde{p}_{1}\,\frac{S(t-\kappa)}{N}c^{U_2}_{\text{pot}}(t) 
	\end{align*}
	and those infected while their $U_2$-index case was in $U_2$
	\begin{align*}
		c^{\text{inf},U_2}_{\text{pot},U_2}(t)= \frac{\abs{J^{U_2,\text{late}}_{t,T}}}{\abs{J^{U_2}_{t,T}}}\,\widetilde{p}_{2}\,\frac{S(t-\kappa)}{N}c^{U_2}_{\text{pot}}(t).
	\end{align*}
	We approximate the time that contacts, infected while their $U_2$-index case was in $U_1$, have spent in the infected chain by the time $t$ of being traced as
	\begin{align}\label{eq:U2_U1_dur}
		\widetilde{r}^{U_2}_{U_1}(t) = \kappa + \frac{1}{2}\abs{J^{U_2,\text{early}}_{t,T}} + \abs{J^{U_2,\text{late}}_{t,T}}.
	\end{align}
	For those contacts, infected while their $U_2$-index case was in $U_2$, we approximate this time as
	\begin{align}\label{eq:U2_U2_dur}
		\widetilde{r}^{U_2}_{U_2}(t) = \kappa + \frac{1}{2}\abs{J^{U_2,\text{late}}_{t,T}}.
	\end{align}
	
	Let us now consider an average index case detected at time $t-\kappa$ while being in $U_1$ ({\it $U_1$-index case}). For simplicity, we assume that the same tracing window $T$ as for $U_2$-index cases is chosen, thus, the index case is asked to disclose close contacts from $J^{U_1}_{t,T}=[t-\kappa-T,t-\kappa]$. Denote the time the average $U_1$-index has been infectious by the time $t-\kappa$ of being detected by $\tau_t^{U_1}$. Depending on the relationship between $T$ and $\tau_t^{U_1}$, the tracing interval $J^{U_1}_{t,T}$ is composed of two subintervals:
	\begin{itemize}
	    \item the potentially trivial part of $J^{U_1}_{t,T}$ during which the index case was not yet infectious
	    $$J^{U_1,\text{ninf}}_{t,T}=\begin{cases}
		\emptyset, &\quad \text{if}\,\, T\leq\tau^{U_1}_t\\
		[t-\kappa-T,t-\kappa-\tau^{U_1}_t], &\quad \text{if}\,\, T>\tau^{U_1}_t
	    \end{cases}$$
	    
	    \item the time in $J^{U_1}_{t,T}$ during which the index case was in the early infectious phase
	    \begin{align}\label{eq:interv_U1}
		J^{U_1,\text{early}}_{t,T}=\begin{cases}
		[t-\kappa-T,t-\kappa], &\quad \text{if}\,\, T\leq\tau^{U_1}_{t}\\
		[t-\kappa-\tau^{U_1}_{t},t-\kappa],&\quad \text{if}\,\, T>\tau^{U_1}_{t}.
	    \end{cases}
	    \end{align}
	\end{itemize}
    We approximate the rate at which contacts of $U_1$-index cases become traceable at time $t$ as
	\begin{align}\label{eq:U1_contacts}
		c^{U_1}_{\text{pot}}(t)=\abs{J^{U_1}_{t,T}}\,\widetilde{c}\,\eta_{U_1}(t-\kappa)U_1(t-\kappa).
	\end{align}
	The rate at which infected contacts of $U_1$-index cases become traceable at time $t$ is approximated as
	\begin{align*}
		c_{\text{pot}}^{\text{inf},U_1}(t)=\frac{\abs{J^{U_1,\text{early}}_{t,T}}}{\abs{J^{U_1}_{t,T}}}\,\widetilde{p}_{1}\,\frac{S(t-\kappa)}{N}c_{\text{pot}}^{U_1}(t). 
	\end{align*}
	The time such contacts have stayed in the infected chain by the time of being traced $t$ is estimated as
	\begin{align}\label{eq:U1_dur}
		\widetilde{r}^{U_1}(t) = \kappa + \frac{1}{2}\abs{J^{U_1,\text{early}}_{t,T}}.
	\end{align}
	The sum of \eqref{eq:U2_contacts} and \eqref{eq:U1_contacts} gives the total rate at which contacts become traceable at time $t$,
	\begin{align}\label{eq:pot_presym}
		c_\text{pot}(t)=c_{\text{pot}}^{U_2}(t) + c_{\text{pot}}^{U_1}(t),
	\end{align}
	which determines the tracing efficiency (see equation \eqref{eq:pNorm})
	\begin{align}
		\varepsilon(t)=\frac{\Omega}{\norm{(c_\text{pot}(t),\Omega)}_p}.
	\end{align}
	The actual rates at which infected contacts are quarantined at time $t$ are then given by
	\begin{align}
			c^{\text{inf},U_1}_{\text{act}}(t)&=c_{\text{pot}}^{\text{inf},U_1}(t)\varepsilon(t), \label{eq:U1_contacts_act}\\
			c^{\text{inf},U_2}_{\text{act},U_1}(t)&=c^{\text{inf},U_2}_{\text{pot},U_1}(t)\varepsilon(t), \label{eq:U2_U1_contacts_act}\\
			c^{\text{inf},U_2}_{\text{act},U_2}(t)&=c^{\text{inf},U_2}_{\text{pot},U_2}(t)\varepsilon(t). \label{eq:U2_U2_contacts_act}
	\end{align}
	In order to approximate from which compartment the infected contacts described by these rates originate, we use the durations \eqref{eq:U1_dur},\eqref{eq:U2_U1_dur},\eqref{eq:U2_U2_dur} and follow the first-order kinetics of system \eqref{eq:fullsys_presym}. To this end, we solve the initial-value problem
	\begin{align*}
	    \frac{d\widetilde{E}}{ds}~&=~-\alpha \widetilde{E}\\
		\frac{d\widetilde{U}_1}{ds}~&=~-(\gamma_1+\eta_{U_1}(t-\kappa))\widetilde{U}_1+\alpha \widetilde{E}\\
		\frac{d\widetilde{U}_2}{ds}~&=~-(\gamma_2+\eta_{U_2}(t-\kappa))\widetilde{U}_2+\gamma_1\widetilde{U}_1\\
		\widetilde{E}(0)~&=~1\\
		\widetilde{U}_1(0)~&=~0\\
		\widetilde{U}_2(0)~&=~0
	\end{align*} 
	and evaluate the solution at time $s=\widetilde{r}^{U_1}(t)$ when considering contacts of $U_1$-index cases; at time $s=\widetilde{r}^{U_2}_{U_1}(t)$ for contacts of $U_2$-index cases that were infected when the respective index case was in $U_1$; and at time $s=\widetilde{r}^{U_2}_{U_2}(t)$ when considering contacts of $U_2$-index cases that were infected when the respective index case was in $U_2$. This results in fractions $\mu^{U_1}_{E}(t)$ of \eqref{eq:U1_contacts_act}, $\mu^{U_2}_{U_1,E}(t)$ of \eqref{eq:U2_U1_contacts_act} and $\mu^{U_2}_{U_2,E}(t)$ of \eqref{eq:U2_U2_contacts_act} originating from the $E$ compartment. Similarly, we get fractions $\mu^{U_1}_{U_1}(t)$ of \eqref{eq:U1_contacts_act}, $\mu^{U_2}_{U_1,U_1}(t)$ of \eqref{eq:U2_U1_contacts_act} and $\mu^{U_2}_{U_2,U_1}(t)$ of \eqref{eq:U2_U2_contacts_act} originating from the $U_1$ compartment and lastly fractions $\mu^{U_1}_{U_2}(t)$ of \eqref{eq:U1_contacts_act}, $\mu^{U_2}_{U_1,U_2}(t)$ of \eqref{eq:U2_U1_contacts_act} and $\mu^{U_2}_{U_2,U_2}(t)$ of \eqref{eq:U2_U2_contacts_act} originating from the $U_2$ compartment.

	We end up with the following terms describing contact tracing
	\begin{align*}
		Tr_E(t) &= \mu^{U_1}_E(t) c^{\text{inf},U_1}_{\text{act}}(t) + \mu^{U_2}_{U_1,E}(t) c^{\text{inf},U_2}_{\text{act},U_1}(t) + \mu^{U_2}_{U_2,E}(t) c^{\text{inf},U_2}_{\text{act},U_2}(t),\\
		Tr_{U_1}(t) &= \mu^{U_1}_{U_1}(t) c^{\text{inf},U_1}_{\text{act}}(t) + \mu^{U_2}_{U_1,U_1}(t) c^{\text{inf},U_2}_{\text{act},U_1}(t) + \mu^{U_2}_{U_2,U_1}(t) c^{\text{inf},U_2}_{\text{act},U_2}(t),\\
		Tr_{U_2}(t) &= \mu^{U_1}_{U_2}(t) c^{\text{inf},U_1}_{\text{act}}(t) + \mu^{U_2}_{U_1,U_2}(t) c^{\text{inf},U_2}_{\text{act},U_1}(t) + \mu^{U_2}_{U_2,U_2}(t) c^{\text{inf},U_2}_{\text{act},U_2}(t).
	\end{align*}
	In the setting of this model we get three different tracing coverages for the three different types of secondary infections. The tracing coverage for individuals who got infected by $U_1$-index cases is given by
	\begin{align}\label{eq:tr_cov_presym1}
		\omega^{U_1}(t):=\frac{\abs{J^{U_1,\text{early}}_{t,T}}}{\tau_t^{U_1}}\frac{\widetilde{\beta}_{1}}{\beta_{U_1}}.
	\end{align} 
	The tracing coverage for individuals who got infected by $U_2$-index cases when these respective index cases were in $U_1$, is given by
	\begin{align}\label{eq:tr_cov_presym2}
		\omega^{U_2}_{U_1}(t):=\frac{\abs{J^{U_2,\text{early}}_{t,T}}}{\tau_t^{U_2,U_1}}\frac{\widetilde{\beta}_{1}}{\beta_{U_1}}.
	\end{align}
	Lastly, the tracing coverage for individuals who got infected by $U_2$-index cases when these respective index cases were in $U_2$, is given by
	\begin{align}\label{eq:tr_cov_presym3}
		\omega^{U_2}_{U_2}(t):=\frac{\abs{J^{U_2,\text{late}}_{t,T}}}{\tau_t^{U_2,U_2}}\frac{\widetilde{\beta}_{2}}{\beta_{U_2}}.
	\end{align}
	We could continue with this scheme and divide the infectious period into more and more subperiods. In this way, we could approximate a realistic infectivity profile along the infectious period. Alternatively, a more elegant approach would be to continuously structure the infected compartment by age of infection \cite{Ferretti2020, Pollmann2021,  Mueller2016, Mueller2000, Fraser2004, Huo2015, Scarabel2021}. However, the approach presented here is sufficient to illustrate the adverse effect of transmission prior to the occurrence of potential symptoms on the effectiveness of TTIQ.

\section{Modeling social and hygiene measures and changes in the tracing coverage}\label{app:measures}
	In this section we discuss how we account for changes in social and hygiene measures (i.e., non constant transmission rates), as well as measures affecting the tracing coverage (e.g., changes in the close contact definition or an increasing awareness to keep track of personal contacts), in the contact tracing terms. As for our derivation of the contact tracing terms in Section 2.2, for simplicity, we explain our approach and assumptions by means of a model without decomposed infectious compartments into early and late infectious phase. We proceeded similarly for the full model \eqref{eq:fullsys_presym} and briefly comment on this case at the end of this section. 
	
	We assume that measures affecting the tracing coverage deviate the contact tracing parameters discontinuously with respect to different "generations" of index cases. Specifically, changes taking place from time $t$ on affect only the contact tracing executed on index case that were detected at time $t-\kappa$ or later, even though these might have overlapping tracing intervals with earlier detected index cases. We indicate this time-dependency of the contact tracing parameters by writing  $\widetilde{p}_t,\widetilde{c}_t,\widetilde{\beta}_t$ for the infection probability, the contact rate and the transmission rate observed by contact tracing, respectively. We refer to this as the contact tracing scheme applied to contacts being traced at time $t$. Accordingly, the rate at which contacts become traceable at time $t$ reads (compare with equation \eqref{eq:pot})
	\begin{align}\label{eq:pot_t}
		c_{\text{pot}}(t)=\abs{J_{t,T}}\,\widetilde{c}_t\,\eta_U(t-\kappa) U(t-\kappa),
	\end{align}
	and the rate at which infected contacts are quarantined at time $t$ is given by (compare with equation \eqref{eq:actinf})
	\begin{align}\label{eq:actinf_t}
		c_{\text{act}}^{\text{inf}}(t)=\abs{J^{\text{inf}}_{t,T}}\,\widetilde{\beta}_t\,\frac{S(t-\kappa)}{N}\eta_U(t-\kappa) U(t-\kappa)\varepsilon(t).
	\end{align}
	
	As discussed in Section \ref{sec:phistar_phibar} changes in the transmission rates are modeled by a time dependent factor $\phi(t)\in[0,1]$, so that 
	$$\beta_U(t)=\phi(t)\overline{\beta_U},$$
	where $\overline{\beta_U}$ is the baseline transmission rate of the disease corresponding to a phase without any intervention. When there is a reduction in transmission rate in place at some time point $t^*$ so that $\beta_U(t^*)=\phi(t^*)\overline{\beta_U}$, this potentially also affects $\widetilde{\beta}_t,\widetilde{p}_t,\widetilde{c}_t$ along their relevant time intervals $J_{t,T}$ and $J^{\text{inf}}_{t,T}$, for all $t$ for which $t^*\in J_{t,T}$. To account for this, we let $\overline{\widetilde{p}_t},\overline{\widetilde{c}_t}$ denote the infection probability and contact rate observed by contact tracing corresponding to a phase without any intervention in combination with the specific contact tracing scheme applied to contacts traced at time $t$. We express deviations from these baseline values due to social and hygiene measures by $\widetilde{\phi}_{1,t}(s)$ and $\widetilde{\phi}_{2,t}(s)$, such that
	\begin{align*}
		\widetilde{c}_t(s)&=\widetilde{\phi}_{1,t}(s)\overline{\widetilde{c}_t},\quad s\in J_{t,T},\\
		\widetilde{p}_t(s)&=\widetilde{\phi}_{2,t}(s)\overline{\widetilde{p}_t},\quad s\in J^{\text{inf}}_{t,T},\\
		\widetilde{\beta}_t(s)&=\widetilde{p}_t(s)\widetilde{c}_t(s),\quad s\in J^{\text{inf}}_{t,T},
	\end{align*}
	whereby given a time $s\in J^{\text{inf}}_{t,T}$ we must have $\widetilde{\beta}_t(s)\leq\beta(s)$. The quantities $\widetilde{\phi}_{1,t}(s)$ and $\widetilde{\phi}_{2,t}(s)$ are influenced by the contact tracing scheme applied at time $t$ indicated by the index $t$ and by the social and hygiene measures that where in place during the time index cases are asked to disclose contacts from, indicated by the argument $s$. Their values could be derived from detailed index case and contact tracing data. However, in this work we confine ourselves to the simplified (non-generic) setting where
	\begin{equation}\label{eq:non-generic}
		\begin{aligned}
			\widetilde{\phi}_{1,t}(s)&=\phi(s),\quad \forall\, t\,\,\text{and}\,\,\forall\,s\in J_{t,T},\\
			\widetilde{\phi}_{2,t}(s)&=1,\quad \forall\, t\,\,\text{and}\,\,\forall\,s\in J^{\text{inf}}_{t,T}.
		\end{aligned}
	\end{equation}
	This resembles that a reduction in general transmission rates by the factor $\phi$ during the tracing interval $J_{t,T}$ reduces contact rates observed by contact tracing at time $t$ by the same factor, independent of the contact tracing scheme applied at time $t$. However, the corresponding infection probabilities are assumed to be unaffected by the particular value of $\phi$.
	
	Equations \eqref{eq:pot_t} and \eqref{eq:actinf_t} require $\widetilde{c}_t$ to be constant on $J_{t,T}$ and $\widetilde{p}_t$ to be constant on $J^{\text{inf}}_{t,T}$ (otherwise there would appear integrals over the respective interval in these equations). Being aware that this does not hold for $\widetilde{c}_t$ when we consider social and hygiene measures as discussed in this section, we assume that $T$ is reasonably small compared to intervals between parameter changes so that we can approximate $\widetilde{c}_t\equiv \widetilde{c}_t(t-\kappa)$ over $J_{t,T}$. This leads to 
	$$\omega(t)=\frac{\abs{J^{\text{inf}}_{t,T}}}{\tau_t}\frac{\widetilde{\beta}_t(t-\kappa)}{\beta_U(t-\kappa)}$$
	being our proxy for the tracing coverage at time $t$. 
	
	The discussed assumptions extend to the model with early and late infectious phase \eqref{eq:fullsys_presym}. Most importantly, we assume that changes to the tracing scheme equally increase or decrease the different tracing coverages \eqref{eq:tr_cov_presym1}-\eqref{eq:tr_cov_presym3}, and that changes in the transmission rates $\beta_{U_2}$, $\beta_{U_1}=\theta\beta_{U_2}$ affect contact rates observed by contact tracing by the same factor but do not affect the corresponding infection probabilities observed by contact tracing.
	
\section{Parameterization}\label{app:parameter}
    Our baseline parameter setting is inspired by the spread of COVID-19 in Germany in late summer and fall of 2020. All scenarios considered in Section 3 in which parameter values deviate from this baseline setting are clearly indicated and explained there.
		
	Parameters describing spreading dynamics and disease characteristics are based on literature dedicated to the original strain of SARS-CoV-2 circulating in 2020. We assume a latent phase of about $1/\alpha=3.5$ days \cite{Li2020}. This latent phase is followed by an early infectious phase of about $1/\gamma_1=2$ days \cite{He2020}, which together with the latent period of about $3.5$ days gives an incubation period of about $5.5$ days for individuals with a symptomatic course of infection \cite{RKI_Steckbrief, Lauer2020, Li2020a}. We assume that individuals go trough an infectious phase of in total about $9$ days and accordingly set $1/\gamma_2=7$ \cite{RKI_Steckbrief, He2020, Bullard2020, Singanayagam2020, Woelfel2020}. The baseline transmission rate in the late infectious compartment is chosen as $\overline{\beta_{U_2}}=0.33$. In addition, we set the scaling factor for the transmission rate of early infectious infectious individuals to $\theta=1.5$. At low prevalence, this leads to approximately $40\%$ \cite{RKI_Steckbrief, Ferretti2020, Ganyani2020, He2020} of transmissions that are produced by undetected infectious individuals to originate from the early phase of infection and to a basic reproduction number of $\mathcal{R}_0=(\theta\overline{\beta_{U_2}})/\gamma_1+\overline{\beta_{U_2}}/\gamma_2=3.3$ \cite{RKI_Steckbrief, Alimohamadi2020, Billah2020, Zhao2020}. 
		
	For the maximal number of tests that can be administered and evaluated per day we only consider polymerase chain reaction tests as rapid antigen tests were not yet widely available in 2020 and assume a baseline value of $\sigma_+=200\,000$. Notice that we do not consider the additional effort generated by testing uninfected traced individuals, the fact that cases might be tested repeatedly or the possibility that all individuals might seek more testing at higher prevalence which would also explain a slower increase in the positive rate. Therefore, we choose a relatively low value for $\sigma_+$ when compared to the theoretical testing capacity reported for Germany in late summer and fall of 2020 \cite{RKI_Tests}. The decay term of tests is set to $\sigma_-=1.353\, N$ such that at low prevalence only approximately $85\,000$ tests are conducted per day. This roughly aligns with the reports for Germany during summer of 2021 at which time the reported incidence was indeed low \cite{RKI_Tests}. It should be noticed, however, that the test capacity in 2021 was significantly increased compared to the considered period of late summer and fall of 2020. Nevertheless, our rough approach captures the detrimental effect of an increasing incidence on the index case detection rates. The relative frequencies $(\sigma_{U_2},\sigma_{Q})$ of testing undetected late infectious and traced individuals compared to susceptibles depend on the considered disease, the testing strategy and the willingness of infectious and suspected individuals to get tested. Here we estimate that $(\sigma_{U_2},\sigma_{Q})=(93,300)$. Individuals in all the other compartments (importantly also those in $U_1$) are assumed to have no increased chance of being tested when compared to a susceptible individual. The choice $\sigma_{U_2}=93$ leads to a conservative case detection ratio of about $40\%$ at low prevalence by testing alone. For simplicity, we choose to work with a constant tracing window $T=1/\gamma_1+1/\gamma_2$, which reflects the theoretical decision to trace contacts of an index case for the average duration of the infectious period. We assume that the tracing scheme leads to a baseline value of $\overline{\widetilde{c}}=0.8$ for the close contact rate reported by an interviewed index case when no contact restrictions are in place. Together with correspondingly selected infection probabilities this results in a tracing coverage of $\omega=0.65$ (below we discuss why do not have to give different values to the different tracing coverages \eqref{eq:tr_cov_presym1}-\eqref{eq:tr_cov_presym3}). The maximal rate at which contacts can be quarantined per day is assumed to be $\Omega=40\,000$. Moreover, we assume a tracing delay of $\kappa=2$ days and a tracing efficiency constant of $p=2$ (a higher $p$ appears to be unrealistic considering the $400$ locally managed public health departments in Germany). The total population size is set to $N=83\,000\,000$ corresponding to the German population as of 2020. 
	
	Our parameter choices are summarized in \autoref{tab:para} in the main text. Due to a lack of detailed contact tracing data for Germany and an unknown case detection ratio, our choices for the TTIQ parameters come with significant uncertainty. However, in the main text sensitivity of our results to these parameters and alternative parameter constellations are investigated. In addition, while we are aware that parameters change much more dynamically in reality, we keep them constant in the considered simulations, apart from single change points. 
	
    \subsubsection*{Tracing terms resulting from the baseline parameter choices}
    Here we specify the exact form of the different components included in the contact tracing terms $Tr_{E}(t)$, $Tr_{U_1}(t)$, $Tr_{U_2}(t)$ in model \eqref{eq:fullsys_presym}  given our above assumptions on parameter choices and the assumptions made in Appendix \ref{app:measures}.
	
	We approximate the periods that the different types of index cases (introduced in Appendix \ref{app:presym}) spent in $U_1$ and $U_2$ before they were identified and isolated at time $t-\kappa$ as
	\begin{align}\label{eq:approx_times}
		\tau_t^{U_1}=\tau_{t,U_1}^{U_2}=\frac{1}{\eta_{U_1}(t-\kappa)+\gamma_1},\quad \tau_{t,U_2}^{U_2}= \frac{1}{\eta_{U_2}(t-\kappa)+\gamma_2},
	\end{align}
	which, considering equations \eqref{eq:interv_U2pre}, \eqref{eq:interv_U2pos} and \eqref{eq:interv_U1}, together with the tracing window $T=1/\gamma_1+1/\gamma_2$, leads to 
	\begin{equation}
		\begin{aligned}\label{eq:approx_ints}
			\abs{J^{U_1}_{t,T}} &= \abs{J^{U_2}_{t,T}} = \frac{1}{\gamma_1}+\frac{1}{\gamma_2},\\
			\abs{J^{U_1,\text{early}}_{t,T}} &= \abs{J^{U_2,\text{early}}_{t,T}} = \frac{1}{\eta_{U_1}(t-\kappa)+\gamma_1},\\
			\abs{J^{U_2,\text{late}}_{t,T}} &= \frac{1}{\eta_{U_2}(t-\kappa)+\gamma_2}.
		\end{aligned}
	\end{equation}
	Using equations \eqref{eq:pot_presym}, \eqref{eq:U1_contacts_act}-\eqref{eq:U2_U2_contacts_act}, \eqref{eq:U1_dur}, \eqref{eq:U2_U1_dur} and \eqref{eq:U2_U2_dur} we get (respecting the assumptions in Appendix \ref{app:measures})
	\begin{align*}
		c_\text{pot}(t) &= \left(\frac{1}{\gamma_1}+\frac{1}{\gamma_2}\right)\,\widetilde{c}_t(t-\kappa)\,\left(\eta_{U_1}(t-\kappa)U_1(t-\kappa)+\eta_{U_2}(t-\kappa)U_2(t-\kappa)\right),\\
		c^{\text{inf},U_1}_{\text{act}}(t) &= \frac{\widetilde{\beta}_{t,1}(t-\kappa)}{\eta_{U_1}(t-\kappa)+\gamma_1}\frac{S(t-\kappa)}{N}\eta_{U_1}(t-\kappa)U_1(t-\kappa)\varepsilon(t),\\
		c^{\text{inf},U_2}_{\text{act},U_1}(t) &= \frac{\widetilde{\beta}_{t,1}(t-\kappa)}{\eta_{U_1}(t-\kappa)+\gamma_1}\frac{S(t-\kappa)}{N}\eta_{U_2}(t-\kappa)U_2(t-\kappa)\varepsilon(t),\\
		c^{\text{inf},U_2}_{\text{act},U_2}(t) &= \frac{\widetilde{\beta}_{t,2}(t-\kappa)}{\eta_{U_2}(t-\kappa)+\gamma_2}\frac{S(t-\kappa)}{N}\eta_{U_2}(t-\kappa)U_2(t-\kappa)\varepsilon(t),\\	
		\widetilde{r}^{U_1}(t) &= \kappa+\frac{1}{2}\frac{1}{\eta_{U_1}(t-\kappa)+\gamma_1},\\
		\widetilde{r}^{U_2}_{U_1}(t) &= \kappa + \frac{1}{2}\frac{1}{\eta_{U_1}(t-\kappa)+\gamma_1} + \frac{1}{\eta_{U_2}(t-\kappa)+\gamma_2},\\
		\widetilde{r}^{U_2}_{U_2}(t) &= \kappa + \frac{1}{2}\frac{1}{\eta_{U_2}(t-\kappa)+\gamma_2}.
	\end{align*}
	Considering equations \eqref{eq:tr_cov_presym1}-\eqref{eq:tr_cov_presym3}, the approximations \eqref{eq:approx_times} and \eqref{eq:approx_ints}, together with assumption \eqref{eq:assum_betas} and the assumptions in Appendix \ref{app:measures}, lead to a single tracing coverage  $$\omega(t):=\omega^{U_1}(t)=\omega^{U_2}_{U_1}(t)=\frac{\widetilde{\beta}_{t,1}(t-\kappa)}{\beta_{U_1}(t-\kappa)}=\frac{\widetilde{\beta}_{t,2}(t-\kappa)}{\beta_{U_2}(t-\kappa)}=\omega^{U_2}_{U_2}(t).$$ 

\section{Stability analysis}\label{app:stability}
	The stability of the DFE is approached by considering the linearization about the DFE of system \eqref{eq:fullsys_presym} \cite{Kuang1993}. The linearization of a system of delay differential equations with a single constant delay takes the form
	\begin{align}\label{eq:linear}
		x'(t)=Ax(t)-Bx(t-\kappa)
	\end{align}
	with matrices $A$ and $B$ depending on the parameters of the considered system. We consider the linearization of model \eqref{eq:fullsys_presym} with respect to the variables $x=(E,Q_E,U_1,Q_{U_1},I_1,U_2,Q_{U_2},I_2)$. It is unnecessary to consider $R$ and $S$ here, since the equation for $R$ is encoded in the remaining state variables and the equation for $S$ only gives a zero eigenvalue since all disease free states are equilibrium solutions. The corresponding matrices are given by
	\[
	A=
	\begin{bmatrix}
		-\alpha & 0 & \beta_{U_1} & p_Q\beta_{U_1} & p_I \beta_{U_1} & \beta_{U_2} & p_Q \beta_{U_2} & p_I \beta_{U_2}\\
		0 & -\alpha & 0 & 0 & 0 & 0 & 0 & 0\\
		\alpha & 0 & -(\overline{\eta_{U_1}}+\gamma_1)  & 0 & 0 & 0 & 0 & 0\\
		0 & \alpha & 0 & -(\overline{\eta_{Q_{U_1}}}+\gamma_1) & 0 & 0 & 0 & 0\\
		0 & 0 & \overline{\eta_{U_1}} & \overline{\eta_{Q_{U_1}}} & -\gamma_1 & 0 & 0 & 0\\
		0 & 0 & \gamma_1 & 0 & 0 & -(\overline{\eta_{U_2}}+\gamma_2) & 0 & 0\\
		0 & 0 & 0 & \gamma_1 & 0 & 0 & -(\overline{\eta_{Q_{U_2}}}+\gamma_2) & 0\\
		0 & 0 & 0 & 0 & \gamma_1 & \overline{\eta_{U_2}} & \overline{\eta_{Q_{U_2}}} & -\gamma_2\\
	\end{bmatrix},
	\]
	\[
	B=
	\begin{bmatrix}
		0 & 0 & -\chi_E^{U_1} & 0 & 0 & -\chi_E^{U_2} & 0 & 0\\
		0 & 0 & \chi_E^{U_1} & 0 & 0 & \chi_E^{U_2} & 0 & 0\\
		0 & 0 & -\chi_{U_1}^{U_1} & 0 & 0 & -\chi_{U_1}^{U_2} & 0 & 0\\
		0 & 0 & \chi_{U_1}^{U_1} & 0 & 0 & \chi_{U_1}^{U_2} & 0 & 0\\
		0 & 0 & 0 & 0 & 0 & 0 & 0 & 0\\
		0 & 0 & -\chi_{U_2}^{U_1} & 0 & 0 & -\chi_{U_2}^{U_2} & 0 & 0 \\
		0 & 0 & \chi_{U_2}^{U_1} & 0 & 0 & \chi_{U_2}^{U_2} & 0 & 0\\
		0 & 0 & 0 & 0 & 0 & 0 & 0 & 0\\
	\end{bmatrix},
	\]
	where
	\begin{align*}
		\chi_E^{U_1} &= \abs{J_{t,T}^{U_1,\text{early}}}\big{|}_{\text{DFE}}\,\,\widetilde{\beta}_{1}\,\,\overline{\eta_{U_1}}\,\,\mu_E^{U_1}\big{|}_{\text{DFE}},\\
		\chi_E^{U_2} &= \abs{J_{t,T}^{U_2,\text{early}}}\big{|}_{\text{DFE}}\,\,\widetilde{\beta}_{1}\,\,\overline{\eta_{U_2}}\,\,\mu_{U_1,E}^{U_2}\big{|}_{\text{DFE}} + \abs{J_{t,T}^{U_2,\text{late}}}\big{|}_{\text{DFE}}\,\,\widetilde{\beta}_{2}\,\,\overline{\eta_{U_2}}\,\,\mu_{U_2,E}^{U_2}\big{|}_{\text{DFE}},\\
		\chi_{U_1}^{U_1} &= \abs{J_{t,T}^{U_1,\text{early}}}\big{|}_{\text{DFE}}\,\,\widetilde{\beta}_{1}\,\,\overline{\eta_{U_1}}\,\,\mu_{U_1}^{U_1}\big{|}_{\text{DFE}},\\
		\chi_{U_1}^{U_2} &= \abs{J_{t,T}^{U_2,\text{early}}}\big{|}_{\text{DFE}}\,\,\widetilde{\beta}_{1}\,\,\overline{\eta_{U_2}}\,\,\mu_{U_1,U_1}^{U_2}\big{|}_{\text{DFE}} + \abs{J_{t,T}^{U_2,\text{late}}}\big{|}_{\text{DFE}}\,\,\widetilde{\beta}_{2}\,\,\overline{\eta_{U_2}}\,\,\mu_{U_2,U_1}^{U_2}\big{|}_{\text{DFE}},\\
		\chi_{U_2}^{U_1} &= \abs{J_{t,T}^{U_1,\text{early}}}\big{|}_{\text{DFE}}\,\,\widetilde{\beta}_{1}\,\,\overline{\eta_{U_1}}\,\,\mu_{U_2}^{U_1}\big{|}_{\text{DFE}},\\
		\chi_{U_2}^{U_2} &= \abs{J_{t,T}^{U_2,\text{early}}}\big{|}_{\text{DFE}}\,\,\widetilde{\beta}_{1}\,\,\overline{\eta_{U_2}}\,\,\mu_{U_1,U_2}^{U_2}\big{|}_{\text{DFE}} + \abs{J_{t,T}^{U_2,\text{late}}}\big{|}_{\text{DFE}}\,\,\widetilde{\beta}_{2}\,\,\overline{\eta_{U_2}}\,\,\mu_{U_2,U_2}^{U_2}\big{|}_{\text{DFE}},\\
	\end{align*}
	and $\overline{\eta_X}$ denotes the detection rate in compartment $X$ evaluated at the disease-free equilibrium.

    The qualitative behavior of solutions of system \eqref{eq:linear} is determined by the roots of the characteristic equation of \eqref{eq:linear} given by
	\begin{align}\label{eq:chara}
		\det(-\lambda I + A + e^{-\lambda \kappa}B)=0.
	\end{align}
	In order to numerically approximate solutions of \eqref{eq:chara} a discretization of the PDE-representation of the DDE can be used. This discretization results in a matrix whose eigenvalues are approximations for solutions of \eqref{eq:chara}. To this end, we followed the discretization scheme based on a Chebyshev nodes described in \cite{Jarlebring2008}. The necessary code presented in \cite{Jarlebring2008} and \cite{Trefethen2000} was -- like all numerical experiments in this work -- implemented in Python \cite{Harris2020, Virtanen2020, Hunter2007, ddeint2019,Vallat2018}. We calculated the critical level $\phi^*$ of effective contacts introduced in Section 2.3 by scanning for the threshold level in effective contacts which determines a stability switch. In other words, for $\phi<\phi^*$ equation \eqref{eq:chara} has only solutions with negative real part and the DFE is stable. For $\phi>\phi^*$ there is at least one solution of \eqref{eq:chara} with positive real part and the DFE is unstable.
\end{appendices}

\end{document}